\documentclass{JHEP3}
\usepackage{amsmath}
\usepackage{cite}
\usepackage{epsfig}
\usepackage{xspace}
\usepackage{graphics}
\usepackage[latin1]{inputenc}

\def\hide#1{}

\newcommand{\as}{\ensuremath{\alpha_{\mathrm{s}}}}

\def\mrm#1{\mathrm{#1}}

\def\f2d3{\ensuremath{F_2^{\mrm{D}3}}}

\def\done#1{}

\providecommand{\eqref}[1]{eq.~(\ref{#1})\xspace}
\renewcommand{\eqref}[1]{eq.~(\ref{#1})\xspace}

\newcounter{aenumct}

\newcounter{ienumct}

\renewenvironment{itemize}{\begin{list}{$\bullet$}%
{\setlength{\topsep}{0mm}\setlength{\partopsep}{1mm}%
\setlength{\itemsep}{0mm}\setlength{\parsep}{1mm}}}%
{\end{list}}

\newcounter{enumct}

\newcommand{\be}[0]{\ensuremath{\beta}}
\newcommand{\si}[0]{\ensuremath{\sigma}}
\newcommand{\te}[0]{\ensuremath{\theta}}

\newcommand{\Ps}[0]{\ensuremath{\Psi}}
\newcommand{\ga}[0]{\ensuremath{\gamma}}

\newcommand{\de}[0]{\ensuremath{\delta}}
\newcommand{\al}[0]{\ensuremath{\alpha}}
\newcommand{\ep}[0]{\ensuremath{\varepsilon}}

\usepackage{axodraw}
\usepackage{epsfig}
\usepackage{amssymb}
\usepackage{mathrsfs}
\usepackage{mathcomp}
\usepackage{cite}
\usepackage[latin1]{inputenc}

\def\pmb#1{{\mbox{\boldmath$#1$}}}

\preprint{LU-TP 10-13\\
\today}

\title{Fluctuations, Saturation, and Diffractive Excitation in High Energy 
Collisions}

\author{Christoffer Flensburg and Gösta Gustafson\\
  Dept.~of Theoretical Physics,
  Sölvegatan 14A, S-223 62  Lund, Sweden\\
  E-mail: \email{christoffer.flensburg@thep.lu.se} 
   and \email{gosta.gustafson@thep.lu.se}}

\abstract{Diffractive excitation is usually described by the Good--Walker 
formalism for low masses, and by the triple-Regge
formalism for high masses. In the Good--Walker formalism the cross section is
determined by the fluctuations in the interaction. In this paper we show that
by taking the fluctuations in the BFKL ladder into account, it is possible to
describe both low and high mass excitation by the Good--Walker mechanism.
In high energy $pp$ collisions the fluctuations are strongly suppressed by 
saturation, which implies that pomeron exchange does not factorise between DIS
and $pp$ collisions. The Dipole Cascade Model reproduces the expected
triple-Regge form for the bare pomeron, and the triple-pomeron coupling is
estimated.}

\begin{document}

\section{Introduction}

Diffractive excitation represents large fractions of the cross sections in
$pp$ collisions or DIS. In most analyses of $pp$ collisions low mass excitation is described by
the Good--Walker formalism \cite{Good:1960ba}, while high mass excitation is
described by a triple-Regge formula \cite{Mueller:1970fa, Detar:1971gn}. 
In the Good--Walker 
formalism the state of the incoming projectile is written as a superposition 
of eigenstates to the $T$-matrix, and the cross section for
diffractive excitation is given by the fluctuations in the eigenvalues. In the
triple-Regge formulation it is instead determined by the reggeon couplings to
the projectile and the target, and a set of triple-reggeon couplings,
determined by fits to data (for recent analyses see e.g. refs. 
\cite{Ryskin:2009tj, Kaidalov:2009aw}). The fluctuations in the pomeron ladder 
are here not included in the Good--Walker formalism, which 
therefore limits the application to low masses. 
It is, however, well known that the fluctuations in the evolution of a BFKL
pomeron are very large \cite{Mueller:1996te}. As we will discuss in the 
following, by including these fluctuations it is possible to describe both low
and high mass diffraction in a uniform way, within the Good--Walker formalism. 

In central $pp$ collisions the interaction is approaching the black
limit at increasing energy, and therefore saturation effects are very important.
The triple-Regge formula would violate unitarity and 
predict a diffractive cross section
exceeding the total cross section if saturation and multiple pomeron 
interactions are not included. These are accounted for in terms of
gap-survival form factors and
``enhanced diagrams'', as in the references cited above, 
or as saturation effects in the pomeron flux \cite{Goulianos:1995vn}.
In the Good--Walker approach
these effects are taken into account by reduced fluctuations,
when the interaction approaches the black limit.

Data from HERA show a very large cross section for diffractive excitation
of the virtual photon. In DIS the photon couples initially to a virtual
$q\bar{q}$ pair. To improve the description of diffractive excitation of the 
photon, gluon radiation has been included. The incoming virtual photon has
been treated as a mixture of $q\bar{q}$ and $q\bar{q}g$ states, and the
data has been fitted to diffractive proton structure functions or
parton distributions in the pomeron
\cite{Nikolaev:1991et, Bartels:1996fu, Bartels:1998ea}. 
In a description in transverse coordinate space
also effects of saturation for small $Q^2$ have been taken into account
\cite{GolecBiernat:2001mm}.
Although important for very small $x$ and small $Q^2$, saturation is much less 
essential in DIS than in $pp$ collisions, which can explain
the lack of factorisation in the comparison of DIS and $pp$ collisions
\cite{Schilling:2002tz} (see e.g. ref.~\cite{Goulianos:2004as}).

The eikonal approximation, formulated in impact parameter space, is a 
formalism which efficiently accounts for
saturation effects and unitarity constraints in high energy reactions.
If the colliding particles have a substructure, the eikonal formalism can 
also describe diffractive excitation within the Good--Walker
formalism. Miettinen and Pumplin \cite{Miettinen:1978jb}
suggested that the scattering eigenstates correspond to
parton showers, which interact via parton-parton scattering. (They also
suggested that the partons might be identical to quarks and gluons,
which at the time were still hypothetical.) The model predicted that 
diffractive excitation is
dominantly peripheral, with a maximum for impact parameter $b\approx
0.5$ fm at $\sqrt s = 53$ GeV.

Mueller and coworkers have developed a dipole cascade model in
transverse coordinate space, which at the same time reproduces leading log
BFKL evolution and satisfies $s$-channel unitarity 
\cite{Mueller:1993rr, Mueller:1994jq, Mueller:1994gb}.
The evolution of the cascade gives dipole chains, which interact via 
gluon exchange. Multiple interactions then correspond 
to the exchange of multiple pomerons.
It was pointed out by Mueller and Salam \cite{Mueller:1996te} that the dipole 
evolution
contains very large fluctuations. This caused a technical problem for their 
MC simulations, but, as discussed below, including the fluctuations 
in the pomeron ladder gives the possibility
to treat also higher mass excitations in the Good--Walker formalism.

In a series of papers \cite{Avsar:2005iz, Avsar:2006jy, Avsar:2007xg, 
Flensburg:2008ag} a generalisation of Mueller's model is presented,
which includes the following improvements:

- NLL BFKL effects

- Nonlinear effects within the evolution

- Confinement effects

- A simple model for the proton wavefunction

This model describes successfully total and (quasi)elastic cross 
sections for DIS and $pp$ collisions. While taking into account not only
fluctuations in the projectile wave function, but in the whole evolution
between the projectile and the target, the model is also able to describe 
diffractive excitation, not only to low, but also to high masses
\cite{Avsar:2007xg}. Studying a collision in
a frame, where the projectile is evolved a distance $Y_{\mathrm{p}}$ in rapidity,
and the target a distance $Y_{\mathrm{t}}= Y-Y_{\mathrm{p}}$, it is possible
to calculate diffractive scattering where the rapidity range of the excited 
projectile, approximately given by $\ln M_{\mathrm{X}}^2$, is smaller than $Y_{\mathrm{p}}$.
(See sec.~\ref{sec:applicationdiff} for details.) Varying 
$Y_{\mathrm{p}}$ then gives the mass distribution 
$d\sigma/d\ln M_{\mathrm{X}}^2 \sim d\sigma/dY_{\mathrm{p}} $.

In a similar way it is possible to calculate double diffractive excitation
for $M_{\mathrm{Xp}}^2<\exp(Y_p)$ and $M_{\mathrm{Xt}}^2<\exp(Y-Y_t)$,
where the projectile and target are excited to $M_{\mathrm{Xp}}$ and $M_{\mathrm{Xt}}$ respectively.
We note that final states, where the two excited states overlap in rapidity, cannot
be calculated in this way; in this formalism they are instead included in the
inelastic cross section. (We want to return to this problem in a future publication.)

The aim of this paper is to study the nature of the fluctuations in the 
evolution of parton cascades in more detail, in order to understand the 
relation between the Good--Walker
and the triple-Regge formalism for diffractive excitation. We will see that
within the dipole cascade model the Good--Walker mechanism indeed reproduces
the expected bare pomeron trajectory and the triple-Regge result for
diffraction. We will also investigate the effects of saturation in more 
detail, and how the absorptive effects and enhanced diagrams correspond to
saturation effects in dipole cascade evolutions, and how this describes the 
breaking of factorisation between DIS and $pp$ scattering.

In the present paper we will not discuss the properties of exclusive final 
states in diffraction, or events with multiple rapidity gaps. We hope to return 
to these questions in future publications. We are also here not 
discussing the nature of hard diffraction, which has been analysed in terms
of a hard parton scattering supplemented with extra gluon exchange
neutralising the colour exchange, together with Sudakov form factors
describing the gap survival probability (see e.g. ref.~\cite{Ryskin:2009tk,Boonekamp:2001vk,Gotsman:2008tr,Cox:1999dw}).

Section 2 of this paper introduces the Good--Walker formalism and how it can be
applied to parton cascades. Section 3 summarises the features of the Lund 
dipole cascade model used in our analysis. The nature of the fluctuations and effects of saturation in
DIS and $pp$ collisions is analysed in section 4, and in section 5  we study
the impact parameter profile and the $t$-dependence in $pp$ scattering. 
The results of the Good--Walker analysis is 
compared to the triple-Regge formalism is section 6, and the bare pomeron 
couplings are estimated. Our conclusions are summarised in section 7.

\section{The eikonal approximation and the Good--Walker formalism}
\label{sec:eikonal}

\subsection{Eikonal approximation}

Diffraction, saturation, and multiple interactions are more easily described in 
impact parameter space. In transverse momentum space the amplitude for 
two successive interactions is represented by a convolution of the single 
interaction contributions, which in impact parameter
space simplifies to a multiplication. 

If the scattering is driven by absorption into a large number of
inelastic states $n$, with Born amplitudes $\sqrt{2f_n}$, the optical
theorem gives an elastic Born amplitude 
\begin{equation}
F=\sum f_n.
\label{eq:F}
\end{equation}
In our notation, where $T\equiv 1-S$, these amplitudes are purely real.
In the eikonal approximation multiple interactions exponentiates,
and the amplitude
\begin{equation}
T = 1-e^{-F}=1-e^{-\sum f_n} \label{eq:unit}
\end{equation}
is always satisfying the unitarity constraint $T\leq1$. For a structureless 
projectile we then find:
\begin{eqnarray}
d\sigma_{\text{tot}} / d^2b& \sim&\,\langle 2T \rangle \nonumber\\
d\sigma_{\text{el}}/d^2b \,\,& \sim& \, \langle T\rangle^2 \nonumber\\
d\sigma_{\text{inel}}/d^2b& \sim&  \,\langle 1-e^{-\sum 2f_n}\rangle = 
d\sigma_{\text{tot}} / d^2b-d\sigma_{\text{el}} / d^2b
\label{eq:eikonalcross}
\end{eqnarray}

\subsection{Good--Walker formalism}

If the projectile has an internal structure, the mass eigenstates can differ
from the eigenstates of diffraction.
We denote the diffractive eigenstates $\Phi_n$, with eigenvalues $T_n$,
and the mass eigenstates $\Psi_{k} = \sum_n  c_{kn} \Phi_n$,
where the incoming state is given by $\Psi_{\text{in}}=\Psi_1$. 

The elastic amplitude is then given by (assuming here that $c_{1n}$ are real)
\begin{equation} 
\langle \Psi_{1} | T | \Psi_{1} \rangle = \sum c_{1n}^2 T_n ,
= \langle T \rangle
\end{equation}
which implies that
\begin{equation}
d \sigma_{\text{el}}/d^2 b = \left( \sum c_{1n}^2 T_n \right)^2 = \langle T \rangle ^2.
\end{equation}
The amplitude for diffractive transition to the mass eigenstate $\Psi_k$
becomes
\begin{equation}
\langle \Psi_{k} | T | \Psi_{1} \rangle = \sum_n  c_{kn} T_n c_{1n},
\end{equation}
which gives a total diffractive cross section (incl. elastic scattering)
\begin{equation}
d\sigma_{\text{diff}}/d^2 b
=\sum_k \langle \Psi_{1} | T | \Psi_{k} \rangle \langle \Psi_{k} | T |
\Psi_{1} \rangle =\langle T^2 \rangle.
\label{eq:GWsigmadiff}
\end{equation}
Subtracting the elastic scattering we find the cross section for 
diffractive excitation
\begin{equation}
d\sigma_{\text{diff ex}}/d^2 b  = d\sigma_{\text{diff}}/d^2 b - d \sigma_{\text{el}}/d^2 b =
\langle T^2 \rangle - \langle T \rangle ^2\equiv V_T,
\label{eq:eikonaldiff}
\end{equation}
which thus is determined by the fluctuations in the scattering
process.

\subsection{What are the diffractive eigenstates?}

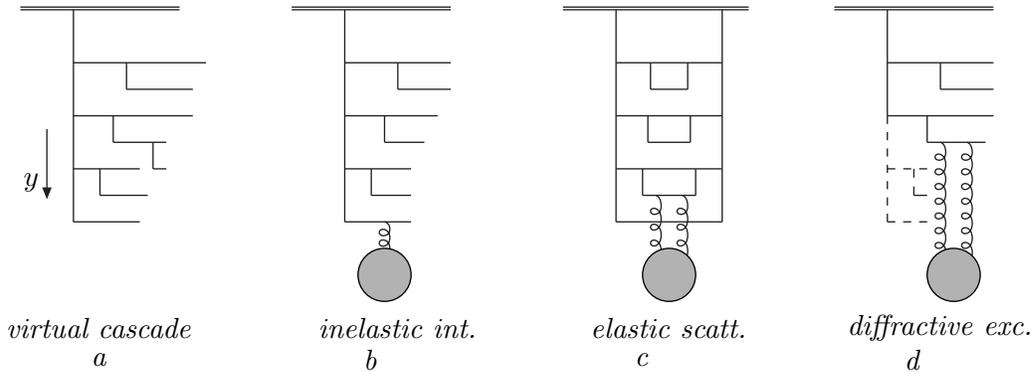
\begin{figure}
\begin{minipage}[]{0.23\linewidth}
\begin{picture}(150,130)(10,-15)

\Line(15,121)(75,121)
\Line(15,120)(75,120)
\Line(35,120)(35,40)
\Line(35,100)(85,100)
\Line(55,100)(55,90)
\Line(55,90)(80,90)
%
\Line(35,80)(80,80)
\Line(50,80)(50,70)
\Line(50,70)(70,70)
\Line(65,70)(65,60)
\Line(65,60)(70,60)
\Line(35,60)(60,60)
\Line(45,60)(45,50)
\Line(45,50)(63,50)
\Line(35,40)(60,40)

\LongArrow(25,75)(25,50)
\Text(22,55)[r]{$y$}
\Text(45,0)[]{\textit{virtual cascade}}
\Text(45,-12)[]{\textit{a}}
\end{picture}
\end{minipage}
\begin{minipage}[]{0.23\linewidth}
\begin{picture}(150,130)(10,-15)

\Line(15,121)(75,121)
\Line(15,120)(75,120)
\Line(35,120)(35,40)
\Line(35,100)(75,100)
\Line(55,100)(55,90)
\Line(55,90)(75,90)
%
\Line(35,80)(70,80)
\Line(50,80)(50,70)
\Line(50,70)(65,70)
\Line(35,60)(60,60)
\Line(45,60)(45,50)
\Line(45,50)(60,50)
\Line(35,40)(60,40)

\Gluon(50,40)(50,20){2}{4}
\GCirc(50,20){10}{0.7}

\Text(55,0)[]{\textit{inelastic int.}}
\Text(45,-12)[]{\textit{b}}

\end{picture}
\end{minipage}
\begin{minipage}[]{0.23\linewidth}
\begin{picture}(150,130)(10,-15)

\Line(15,121)(85,121)
\Line(15,120)(85,120)
\Line(35,120)(35,40)
\Line(75,120)(75,40)
\Line(35,100)(75,100)
\Line(48,100)(48,90)
\Line(62,100)(62,90)
\Line(48,90)(62,90)
%
\Line(35,80)(75,80)
\Line(47,80)(47,70)
\Line(63,80)(63,70)
\Line(47,70)(63,70)
\Line(35,60)(75,60)
\Line(45,60)(45,50)
\Line(65,60)(65,50)
\Line(45,50)(65,50)
\Line(35,40)(75,40)

\Gluon(50,50)(50,20){2}{4}
\Gluon(60,50)(60,20){2}{4}
\GCirc(55,20){10}{0.7}

\Text(55,0)[]{\textit{elastic scatt.}}
\Text(45,-12)[]{\textit{c}}

\end{picture}
\end{minipage}
\begin{minipage}[]{0.25\linewidth}
\begin{picture}(150,130)(10,-15)

\Line(15,121)(75,121)
\Line(15,120)(75,120)
\Line(35,120)(35,80)
\DashLine(35,80)(35,40){3}
\Line(35,100)(75,100)
\Line(55,100)(55,90)
\Line(55,90)(75,90)
%
\Line(35,80)(75,80)
\Line(50,80)(50,70)
\Line(50,70)(72,70)
\DashLine(35,60)(50,60){3}
\DashLine(45,60)(45,50){3}
\DashLine(45,50)(50,50){3}
\DashLine(35,40)(50,40){3}

\Gluon(55,70)(55,20){2}{8}
\Gluon(65,70)(65,20){2}{8}
\GCirc(60,20){10}{0.7}
\Text(55,0)[]{\textit{diffractive exc.}}
\Text(45,-12)[]{\textit{d}}
\end{picture}
\end{minipage}
\caption{\label{fig:eigenstates} (a) An example of a parton (or dipole) 
  cascade evolved in
  rapidity. (b) The exchange of a gluon gives rise to an inelastic
  interaction. (c) Elastic scattering is obtained from coherent scattering of
  different partons in different cascades, via the exchange of two gluons. (d) 
  Diffractive excitation is obtained when the result of the two-gluon exchange
  does not correspond to the coherent initial proton state. Here the dashed
  lines indicate virtual emissions, which are not present in the diffractive
  final state.}
\end{figure}

As mentioned in the introduction, Miettinen and Pumplin \cite{Miettinen:1978jb}
assumed that the diffractive eigenstates correspond to
parton cascades, which can come on shell through interaction with the target.
This was also the assumption in our earlier analysis of diffractive
excitation in \cite{Avsar:2007xg}.
The process is illustrated in fig. \ref{fig:eigenstates}. Fig. $a$ shows the
virtual cascade before the collision, and fig. $b$ illustrates an inelastic
interaction, where gluon exchange gives a colour connection between
the projectile and the target. Fig. $c$ shows an elastic interaction, where
two gluons scatter coherently on the partons in the projectile cascade.
It is obtained from the projection of the scattered state onto the
incoming mixture of different cascades. Fig. $d$, finally, shows the 
contribution of the scattered state, which is orthogonal to the incoming
state, and thus corresponds to diffractive excitation. The lines can 
symbolise gluons in a traditional cascade, or dipoles in a dipole cascade.
In fig.~$d$ the dashed lines corresponds to virtual emissions in the cascade,
which cannot come on shell via momentum exchange from the exchanged gluon pair.

A similar approach was also used by Hatta \textit{et al.} 
\cite{Hatta:2006hs}. Their analysis was, however, limited to relatively low
mass excitations. As the authors sought an analytic solution, they studied
very high energies, where the fluctuations in the pomeron evolution 
could be neglected due to saturation. Thus only fluctuations coming from
ordered DGLAP chains close to the virtual photon end of the process were 
included, and therefore it was not possible to treat excitation to 
larger masses.

\section{The dipole cascade model}

\subsection{Mueller's dipole model}

Mueller's dipole cascade model 
\cite{Mueller:1993rr,Mueller:1994jq,Mueller:1994gb} is a formulation
of BFKL evolution in transverse coordinate space. 
Gluon radiation from the colour charge in a parent quark or gluon is screened 
by the accompanying anticharge 
in the colour dipole. This suppresses emissions at large transverse separation,
which corresponds to the suppression of small $k_\perp$ in BFKL.
For a dipole $(\pmb{x},\pmb{y})$ the probability per unit rapidity ($Y$) for
emission of a gluon at transverse position $\pmb{z}$ is given by
\begin{eqnarray}
\frac{d\mathcal{P}}{dY}=\frac{\bar{\alpha}}{2\pi}d^2\pmb{z}
\frac{(\pmb{x}-\pmb{y})^2}{(\pmb{x}-\pmb{z})^2 (\pmb{z}-\pmb{y})^2},
\,\,\,\,\,\,\, \mathrm{with}\,\,\, \bar{\alpha} = \frac{3\alpha_s}{\pi}.
\label{eq:dipkernel1}
\end{eqnarray}
This emission implies that the dipole is split into two dipoles, which
(in the large $N_c$ limit) emit new gluons independently. The result is a
cascade, where the number of dipoles grows exponentially with $Y$.

In a high energy collision, the dipole cascades in the projectile and the 
target are evolved from their rest frames to the rapidities they will have
in the specific Lorentz frame chosen for the analysis.
The growth in the number of dipoles also implies a strong growth for the 
scattering probability, which, however, is kept below 1 by the possibility
to have multiple dipole interactions in a single event.
The scattering probability between two
elementary colour dipoles with coordinates $(\pmb{x}_i,\pmb{y}_i)$ and
$(\pmb{x}_j,\pmb{y}_j)$ in the projectile and the target respectively, is 
given by $2f_{ij}$, where (in Born approximation)
\begin{equation}
  f_{ij} = f(\pmb{x}_i,\pmb{y}_i|\pmb{x}_j,\pmb{y}_j) =
  \frac{\as^2}{8}\biggl[\log\biggl(\frac{(\pmb{x}_i-\pmb{y}_j)^2
    (\pmb{y}_i-\pmb{x}_j)^2}
  {(\pmb{x}_i-\pmb{x}_j)^2(\pmb{y}_i-\pmb{y}_j)^2}\biggr)\biggr]^2.
\label{eq:dipamp}
\end{equation}
The optical theorem then implies that the elastic amplitude for dipole $i$ 
scattering off dipole $j$ is 
given by $f_{ij}$. Summing over $i$ and $j$ gives the one-pomeron elastic 
amplitude
\begin{equation}
F=\sum f_{ij}.
\end{equation}
In the eikonal approximation
the unitarised amplitude is given by the exponentiated expression
\begin{equation}
T(\pmb{b})=1-e^{-F},
\label{tf-relationmueller}
\end{equation}  
and the total, diffractive, and elastic cross sections are given by
the expressions in eqs. (\ref{eq:eikonalcross}, \ref{eq:eikonaldiff}).

\subsection{The Lund dipole cascade model}

In refs.~\cite{Avsar:2005iz, Avsar:2006jy, Flensburg:2008ag} we describe a 
modification of Mueller's cascade model with the following features:
\begin{itemize}
\item It includes essential NLL BFKL effects.
\item It includes non-linear effects in the evolution.
\item It includes effects of confinement.
\end{itemize}

The model also includes a simple model for the proton wavefunction, and is 
implemented in a Monte Carlo simulation program called DIPSY.
Here the NLL effects significantly reduce the production of small dipoles,
and thereby also the associated numerical difficulties with very large 
dipole multiplicities are avoided. As discussed in the cited references,
the model is able to describe a wide range of observables in DIS and
$pp$ scattering, with very few parameters.

\subsubsection{NLL effects}

The NLL corrections to BFKL evolution have three major sources
\cite{Salam:1999cn}:
\vspace{2mm}

\emph {The running coupling:}

This is relatively easily included in a MC simulation process.
\vspace{2mm}

\emph {Non-singular terms in the splitting function:}

These terms suppress large $z$-values in the individual parton branchings, 
and prevent the daughter from being faster than her recoiling parent. 
Most of this effect is taken care of by including energy-momentum conservation
in the evolution. This is effectively taken into account by associating a
dipole with transverse size $r$ with a transverse momentum $k_\perp = 1/r$,
and demanding conservation of the lightcone momentum $p_+$ in every step
in the evolution. This gives an effective cutoff for small dipoles,
which eliminates the numerical problems encountered in the MC
implementation by Mueller and Salam \cite{Mueller:1996te}.
\vspace{2mm}

\emph {Projectile-target symmetry:}

This is also called energy scale terms, and is essentially equivalent
to the so called consistency constraint. This effect is taken into account by 
conservation of both positive and negative lightcone momentum components,
$p_+$ and $p_-$.
The treatment of these effects includes also effects beyond NLL, in a way 
similar to the treatment by Salam in ref.~\cite{Salam:1999cn}. Thus the power 
$\lambda_{\mathrm{eff}}$,
determining the growth for small $x$, is not negative for large
values of $\alpha_s$.

\subsubsection{Non-linear effects and saturation}

As mentioned above, dipole loops (or equivalently pomeron loops) are not
included in Mueller's cascade model,
if they occur within the evolution, but only if they are cut in the Lorentz
frame used in the calculations, as a result of multiple scattering in this 
frame. The result is therefore not frame independent. 
(The situation is similar in the Colour Glass Condensate or the 
JIMWLK equations.) As for dipole scattering
the probability for such loops is given by $\alpha_s$, and therefore
formally colour suppressed compared to dipole splitting, which is
proportional to $\bar{\alpha}=N_c \alpha_s/\pi$. These loops are therefore
related to the probability that two dipoles have the same colour. Two dipoles
with the same colour form a quadrupole. Such a field may be better
approximated by two dipoles formed by the closest colour-anticolour
charges. This corresponds to a recoupling of the colour dipole chains. We call this process a dipole ``swing''. The swing gives rise to loops within the cascades, and makes the cross section frame independent up to a few percent.
We note that a similar effect would also be obtained from gluon exchange 
between the two dipoles.

In the MC implementation each dipole is assigned one of $N_C^2$ colours, 
and dipoles with the same colour are allowed to recouple. The weight for the recoupling is assumed to be 
proportional to $r_1^2 r_2^2/(r_3^2 r_4^2)$,
where $r_1$ and $r_2$ are the sizes of the original dipoles and $r_3$ 
and $r_4$ are the sizes of the recoupled dipoles.
We note that in this formulation the number of dipoles is not reduced.
The given weight favours the formation of smaller 
dipoles, and the saturation effect is obtained because the
smaller dipoles have smaller cross sections. Thus in an evolution
in momentum space the swing would not correspond to an absorption of gluons 
below the saturation line $k_\perp^2 = Q_s^2(x)$; it would rather correspond 
to lifting the gluons to higher $k_\perp$ above this line.

Although this mechanism does not give an explicitly frame independent result, 
MC simulations show that it is a very good approximation.

\subsubsection{Confinement effects}

Confinement effects are included via an effective gluon mass, which
gives an exponential suppression for very large dipoles \cite{Avsar:2007xg}. 
This prevents the proton to grow too fast in transverse size, and is also
essential to satisfy Froisart's bound at high energies \cite{Avsar:2008dn}.

\subsubsection{Initial dipole configurations}

{\bf Photon wavefunction}

An initial photon is split into a $q\bar{q}$ pair, and
for larger $Q^2$ the wavefunction for a virtual photon can be determined 
perturbatively. The well known result has the following form:
\begin{eqnarray}
  \Ps^{\ga 0}_{f h \bar{h}}(Q,r,z) &=& \frac{\sqrt{ \al_{EM} N_C}}{\pi} e_f 
  Q z (1-z)K_0 (r \ep_f ) \de_{h \bar{h}}  \nonumber\\
  \Ps^{\ga +}_{f h \bar{h}}(Q,r,z) &=&
  \frac{\sqrt{ \al_{EM} N_C / 2}}{\pi} e_f \\
  &\times&
  \Big\{ ie^{i\te} \left(  z \de_{h+} \de_{\bar{h}-} - (1-z) \de_{h-} \de_{\bar{h}+}
    \right) \ep_f K_1 (r \ep_f ) 
+ \de_{h+} \de_{\bar{h}+} m_f K_0 (r \ep_f) \Big\} \nonumber
 \label{eq:photonWF}
\end{eqnarray}
with
\begin{equation}
\ep_f = \sqrt{z(1-z)Q^2 + m_f^2} .
\end{equation}
Here $r$ is the transverse size of the dipole, and $z$ is the energy fraction
carried by the quark,
$\lambda = 0$ and $+$ denote the longitudinal and transverse
wavefunctions respectively, $f$ denotes the quark flavour, and $K_0$
and $K_1$ are modified Bessel functions. $e_f$ is the electric charge
of the quark in units of the proton charge and $m_f$ the effective
mass of the quark. For smaller $Q^2$ a hadronic component has to be added,
as described in more detail in ref.~\cite{Flensburg:2008ag}.

{\bf Proton wavefunction}

The internal structure of the proton is governed by soft QCD, and is not 
possible to calculate perturbatively. In the our model it is represented by
an equilateral triangle formed by three dipoles, and with a radius of
$3$ GeV$^{-1} \approx 0.6$ fm. The model should be used at low $x$,
and when the system is evolved over a large 
rapidity range the observable results depend only weakly on the exact 
configuration of the dipoles, or whether the charges are treated as 
(anti)quarks or gluons.

\subsection{Application to diffraction}
\label{sec:applicationdiff}

\begin{figure}
 \scalebox{1.3}{\mbox{
\begin{picture}(150,130)(-60,0)
\Text(20,132)[bl]{{\small \text {proj.}}}
\Line(20,130)(80,130)
\Line(20,131)(80,131)
\Line(40,130)(40,70)
\Line(60,130)(60,70)
\Line(40,115)(60,115)
\Line(40,100)(60,100)
\Line(40,85)(60,85)
\Line(40,70)(60,70)

\Text(90,100)[l]{{\large $\Psi_{\text{proj}}=\sum_n c_n \Phi_{\text{p},n}$}}

\Line(33,55)(12,10)
\Line(33,55)(40,70)
\Line(42,55)(21,10)
\Line(42,55)(48,70)
\Line(58,55)(79,10)
\Line(52,70)(58,55)
\Line(67,55)(88,10)
\Line(60,70)(67,55)

\Line(33,55)(42,55)
\Line(58,55)(67,55)
\Line(26,40)(35,40)
\Line(65,40)(74,40)
\Line(19,25)(28,25)
\Line(72,25)(81,25)
\Line(2,10)(31,10)
\Line(2,9)(31,9)
\Line(69,10)(98,10)
\Line(69,9)(98,9)
\Text(2,7)[tl]{{\small \text {target}}}

\Text(90,40)[l]{{\large $\Psi_{\text{target}}=\sum_m d_m \Phi_{\text{t},m}$}}

\Line(-15,131)(-5,131)

\Line(-15,9)(-5,9)
\LongArrow(-10,110)(-10,130)
\LongArrow(-10,90)(-10,65)
\LongArrow(-10,50)(-10,65)
\LongArrow(-10,30)(-10,10)
\Text(-10,100)[]{$Y_{\mathrm{p}}$}
\Text(-10,40)[]{$Y_{\mathrm{t}}$}

\DashLine(-15,65)(25,65){4}
\DashLine(50,137)(50,2){7}
\end{picture}
}}
\caption{\label{fig:tripleregge}Single diffractive excitation with no final state particles in the $Y_{\mathrm{t}}$ range. The virtual target evolutions are summed on amplitude level, while the real projectile evolutions are summed on cross section level.}
\end{figure}
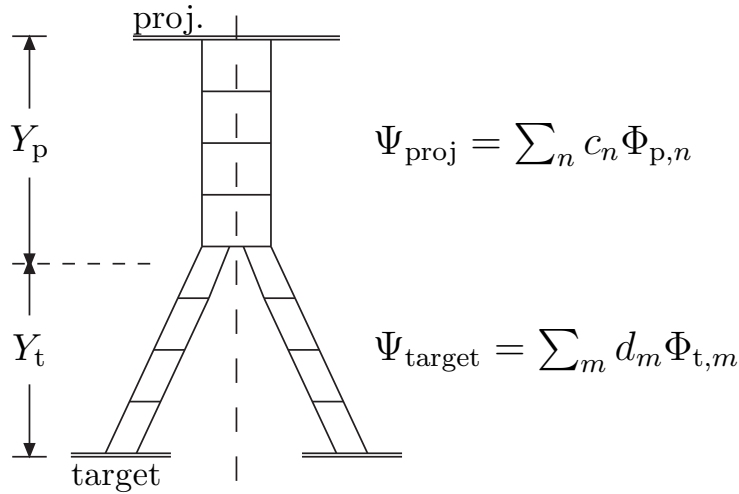

We now want to apply the Good--Walker result in eq.~(\ref{eq:eikonaldiff})
to the situation where two different cascades collide.
The elastic scattering amplitude is obtained when $T$ is averaged over
both the projectile and the target states, while the total diffractive
cross section is obtained by averaging $T^2$. Thus we have
\begin{eqnarray}
d\sigma_{\text{el}}/d^2b = \langle T\rangle_{\mathrm{pt}}^2 \label{eq:sigmael}\\
d\sigma_{\text{diff}}/d^2 b=\langle T^2 \rangle_{\mathrm{pt}}
\end{eqnarray}
Here the indices p and t indicate averaging over the projectile and
target evolutions respectively.
If we average the amplitude over possible evolutions of the target
system, we get the amplitude representing elastic scattering of the target,
If we then square, and average over projectile states, we get according
to eq.~(\ref{eq:GWsigmadiff}), the cross section for total diffractive 
scattering of the projectile, while the target is only scattered elastically.
Subtracting the cross section for elastic scattering of both the projectile
and the target gives the cross section for single diffractive excitation
of the projectile:
\begin{equation}
d\sigma_{\text{proj diff ex}}/d^2 b=\langle \langle T\rangle_{\mathrm{t}}^2 \rangle_{\mathrm{p}}
-\langle T\rangle_{\mathrm{pt}}^2
\label{eq:sigmadiff}
\end{equation}
The process is illustrated in fig.~\ref{fig:tripleregge}.
If the expression is calculated in a Lorentz frame in which the projectile
is evolved a rapidity range $Y_{\mathrm{p}}$, the partons in the projectile
cascade are confined to the rapidity range $y<Y_{\mathrm{p}}$. The
result in eq.~(\ref{eq:sigmadiff}) includes all cascades limited to this
range, also those which have no partons close to $Y_{\mathrm{p}}$. 
This corresponds to all possible
excitation masses $M_{\mathrm{X}}^2 \leq \exp(Y_{\mathrm{p}}) \cdot 1 \mathrm{GeV}^2$. By varying 
$Y_{\mathrm{p}}$ it is then possible to calculate the differential cross section
$d\sigma_{\text{diff ex}}/dM_{\mathrm{X}}^2$.
Final states with $M_{\mathrm{X}}^2 > \exp(Y_{\mathrm{p}}) \cdot 1 \mathrm{GeV}^2$
are thus not included in the cross section in eq.~(\ref{eq:sigmadiff}), in the frame chosen for the
calculation. These states are in our formalism instead included in the inelastic cross section,
because in such a frame there is colour exchange connecting the forward- and backward-moving
systems. To get the full cross section for single diffractive excitation of the projectile,
we must do the calculation in the target rest frame, where $Y_p=Y\equiv \ln s$.
In the same way it is possible to calculate single excitation of the target,
by replacing the role of projectile and target.

The cross section for diffractive scattering of both the projectile and the
target is obtained by $\langle  T^2 \rangle_{\mathrm{pt}}$. This expression
includes both elastic scattering and single diffractive excitation of the
projectile or the target. Subtracting these contributions using 
eqs.~(\ref{eq:sigmael}) and (\ref{eq:sigmadiff}), we get the cross section for
double diffractive excitation given by
\begin{equation}
d\sigma_{\text{DD}}/d^2 b=\langle  T^2 \rangle_{\mathrm{pt}}-
\langle \langle T\rangle_{\mathrm{t}}^2 \rangle_{\mathrm{p}}-
\langle \langle T\rangle_{\mathrm{p}}^2 \rangle_{\mathrm{t}}+
\langle T\rangle_{\mathrm{pt}}^2.
\end{equation}
This expression gives the cross section for 
$M_{\mathrm{Xp}}^2 \leq \exp(Y_{\mathrm{p}})\cdot 1 $GeV$^2$ and 
$M_{\mathrm{Xt}}^2 \leq \exp(Y_{\mathrm{t}})\cdot 1 $GeV$^2$, where 
$Y_{\mathrm{p}}+Y_{\mathrm{t}}$ equals the total rapidity range $Y$.
As was the case for single diffractive excitation, events with excitation
to larger masses are in this formalism included in the inelastic cross
section. For single diffraction it was possible to include excitation of e.g.
the projectile to all masses by performing the calculation in the target rest
frame. This is not the case for double diffraction. Even if we change Lorentz
frame, we can never include events where the two excited states overlap
in rapidity. Those states will always be included in the inelastic cross
section. (Thus although the total and elastic cross sections have to
be independent of the Lorentz frame used, only the sum of the cross sections
for inelastic scattering and diffractive excitation is frame independent.)

The results from MC simulations of single and double diffractive excitation
were presented in ref.~\cite{Avsar:2007xg}, in good agreement with data
from HERA and the Tevatron. Fig.~\ref{fig:sigmatot} shows the diffractive cross sections
for $pp$ collisions at 1800 GeV. In this figure
the projectile is evolved over $Y_\mathrm{p}$ units of rapidity, and the target over 
$Y_\mathrm{t} = Y - Y_\mathrm{p}$ units, setting the limits for the diffracted masses to $M_{\mathrm{Xp}}^2 \leq \exp(Y_\mathrm{p})\cdot 1 $GeV$^2$ and, in case of double diffraction, $M_{\mathrm{Xt}}^2 \leq \exp(Y-Y_\mathrm{p})\cdot 1 $GeV$^2$.

\FIGURE[t]{\includegraphics[angle=270, scale=1.0]{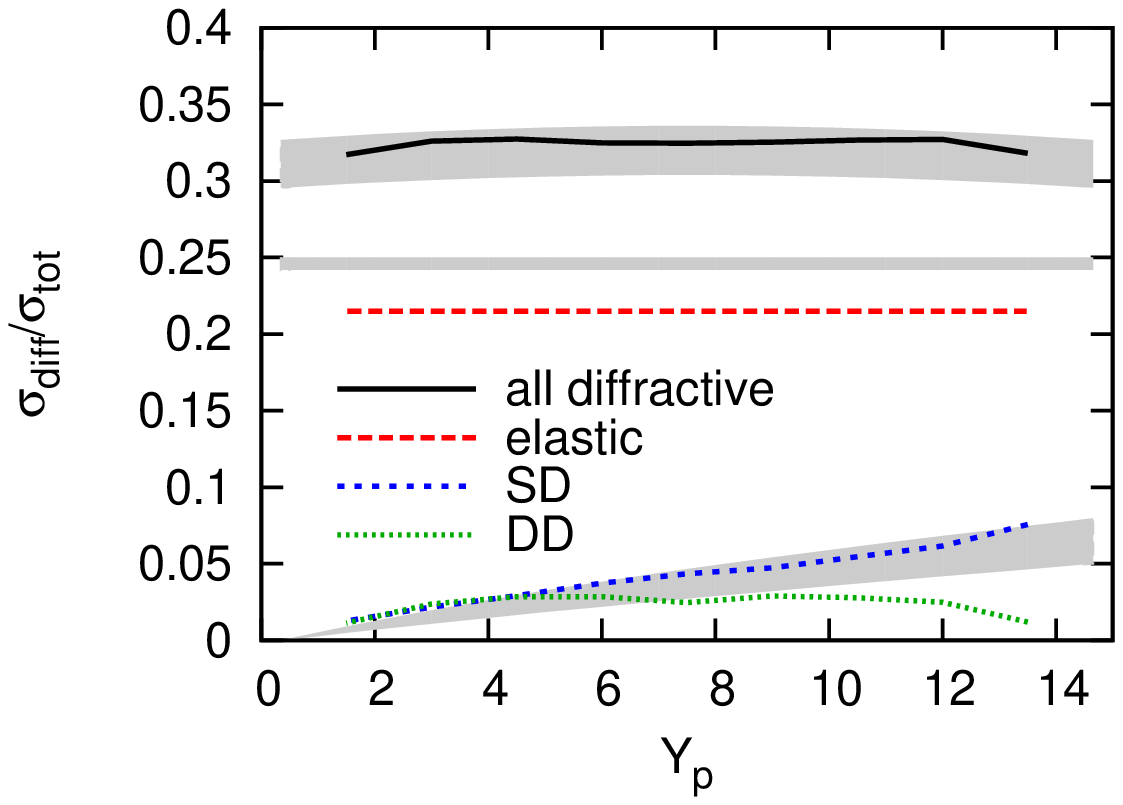}
   \caption{\label{fig:sigmatot}The fraction of elastic, single diffractive and double diffractive events at 1800 GeV as function of interaction frame. For single diffractive excitation the figure shows masses of 
 the excited projectile integrated over $M_{\mathrm{Xp}}^2 \leq \exp(Y_p)\cdot 1 $GeV$^2$.
For double diffraction the projectile and target masses are integrated over $M_{\mathrm{Xp}}^2 \leq \exp(Y_p)\cdot 1 $GeV$^2$ and $M_{\mathrm{Xt}}^2 \leq \exp(Y_t)\cdot 1 $GeV$^2$ respectively, with $Y_p + Y_t =Y= \ln(1800^2) \approx 15$. (For details see the main text.) The two lower error bands are single diffractive excitation and elastic cross section estimated from CDF data \cite{Abe:1993wu,Abe:1993xy}. The top area is the sum of the two, thus not including double diffraction.}} 

We will in the next two sections study how the results follow from the nature of the fluctuations causing
the excitations, and how the fluctuations are suppressed by saturation effects.
We also note that in this approach the effective triple-pomeron coupling is fixed
by the constraint, that it is the same dynamics that determines both the coupling
between the three pomeron ladders in fig.~\ref{fig:tripleregge}, and the
evolution within the individual ladders. The relation to the triple-Regge formalism will be discussed in
sec.~6.

\section{The nature of the fluctuations and effects of saturation}

\subsection{$\gamma^* p$ scattering}
 
The photon wavefunction in eq.~(\ref{eq:photonWF})
is divergent for small dipole sizes, which means that infinitely many small 
dipoles are created with infinitely small cross sections. To illustrate the
fluctuations in the dipole cascade we show in fig.~\ref{fig:f-distDIS}
MC results for the probability distribution, $P(F)$, for the one pomeron 
amplitude $F$ in eq.~(\ref{eq:F}) for a dipole
with a fixed size $r=1/Q$ at a fixed impact parameter $b$. The distribution $P(F)$ 
is here defined so that $P(F) dF$ is the
probability for the formation of a pair of a projectile and a target cascade,
for which the Born amplitude $F=\sum f_{ij}$ lies between $F$ and $F+dF$.
The calculations are performed in the hadronic 
cms, which implies that the diffractive masses are integrated over the range
$M_X^2<\sqrt{W^2}\cdot 1\,\text{GeV}$. 

As seen in fig.~\ref{fig:f-distDIS}, the probability distributions can for all $b$-values
be well approximated by a power spectrum
\begin{equation}
P(F) \approx A\, F^{-p},
\label{eq:powerdistr}
\end{equation}
with a cutoff for small $F$-values. These approximations are shown by the
dotted lines. The two parameters $A$ and $p$ are tuned to fit the MC results for different values of the energy $W$, dipole size $1/Q$, and impact parameter $b$. (The
cutoff is then adjusted to satisfy the normalization condition $\int P(F) dF=1$.) As we will see below, it is particularly interesting to note, that the
fitted value for the power $p$ is independent of the impact parameter $b$. It
varies, however, slowly with $Q^2$ and $W$ as can be seen in table \ref{tab:PFpar}.

\FIGURE[t] {\includegraphics[angle=270, scale=0.59]{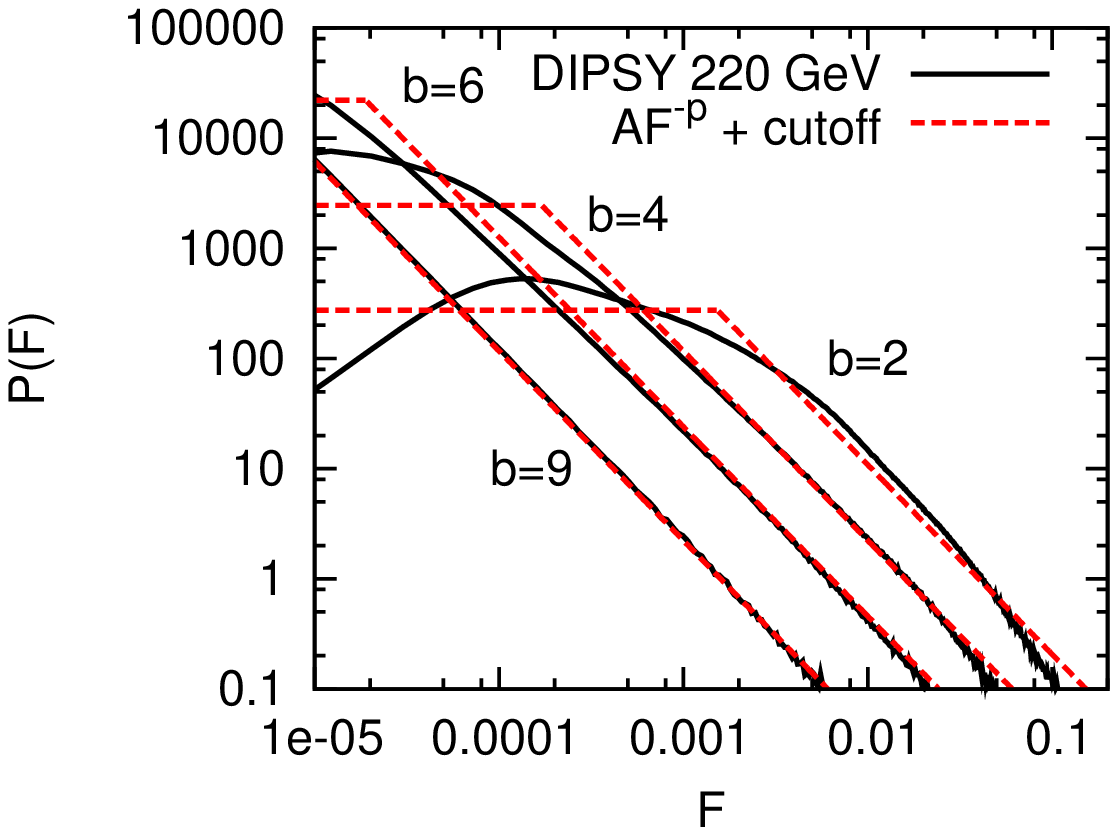}
\includegraphics[angle=270, scale=0.59]{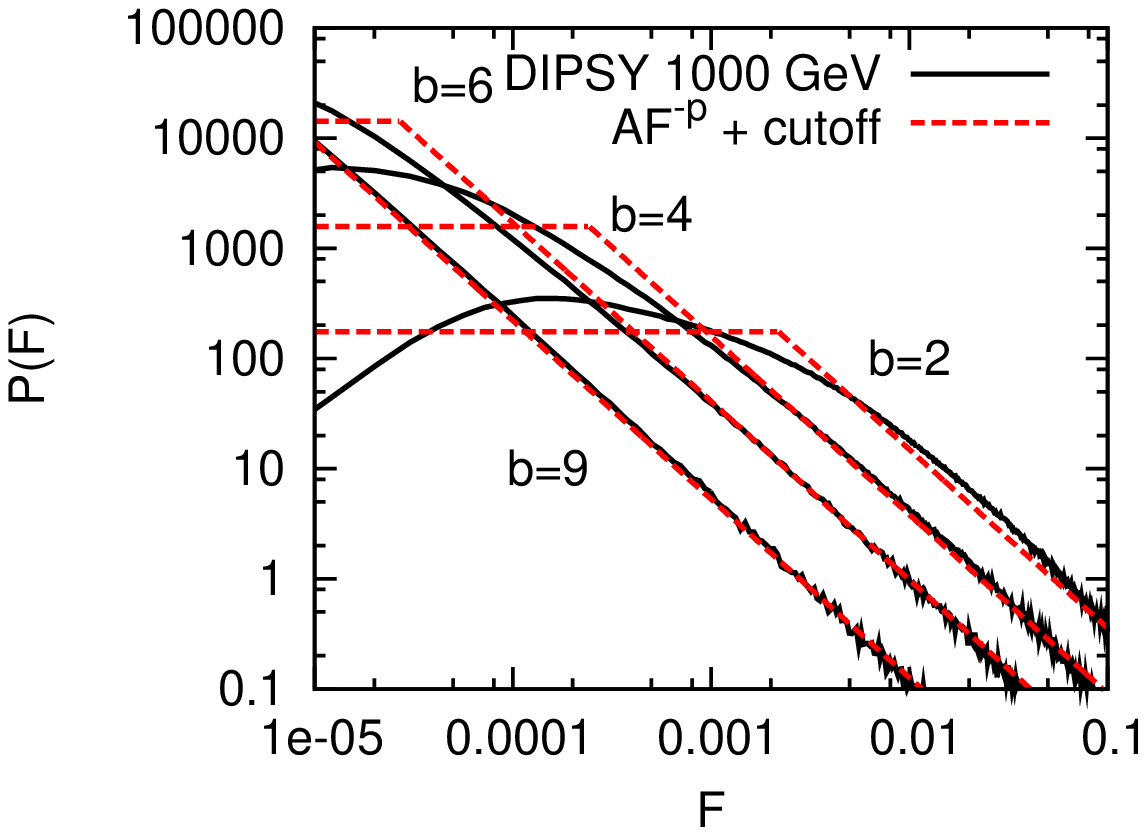}
   \caption{\label{fig:f-distDIS}Probability distribution, $P(F)$, for the one-pomeron
amplitude $F$ in DIS, represented by a dipole with size $r=1/Q$, 
for $Q^2=14$ $\mathrm{GeV}^2$ and $W=220$ $\mathrm{GeV}$
(left) and $W=1000$ $\mathrm{GeV}$ (right). $b$ is in units of GeV$^{-1}$. The dotted lines are fits of the form in eq. (\ref{eq:powerdistr}). } }

\TABLE[t] {
\begin{minipage}{0.49\linewidth}
  $\gamma^\star p$\\[1mm]
  \begin{tabular}{|c||c|c|c|c| }
    \hline
    $W/$GeV  & 220 & 220 & 1000 & 1000  \\
    $Q^2/$GeV$^2$  & 14 & 50 & 14 & 50  \\
    \hline
    $p$           & 1.7 & 1.8  & 1.6 & 1.7 \\
    \hline
  \end{tabular}
\end{minipage}
\begin{minipage}{0.49\linewidth}
  $pp$\\[1mm]
  \begin{tabular}{|c||c|c|c|c| }
    \hline
    $W/$GeV  & 100 & 100 & 2000 & 2000  \\
    $b \cdot$GeV  & 0 & 6 & 0 & 6  \\
    \hline
    $a$           & 1.4 & 1.4 & 0.8 & 0.8 \\
    $p$           & 1.2 & -0.7 & 1.5 & -0.5 \\
    \hline
  \end{tabular}
\end{minipage}
  \caption {\label{tab:PFpar} The values of the parameters in the fits to $P(F)$ for $\gamma^\star p$, eq. (\ref{eq:powerdistr}), and $pp$, eq. (\ref{eq:gammadistr}), for some sample energies and $b$-values. The power $p$ is independent of impact parameter in $\gamma^\star p$, while for $pp$ it is the exponential suppression $a$ that does not depend on $b$.}
}

The cross sections obtained from these distributions can most easily
be estimated from the approximation in eq.~(\ref{eq:powerdistr}).
We see in fig.~\ref{fig:f-distDIS} that the Born amplitudes  are generally small, which
implies that unitarity effects are small, and $T = 1 - e^{-F} \approx F$. 
We also note that the widths of the distributions are large, which means
that $\langle T \rangle ^2$ can be neglected compared to $ \langle T^2 \rangle$. The approximation in eq. (\ref{eq:powerdistr}) then gives the result
\begin{eqnarray}
\frac{d\sigma_{\text{tot}}}{d^2 b}=
2\langle T \rangle& =&2A\int_0^\infty (1-e^{-F})F^{-p} dF = -2A\,\Gamma(1-p);\nonumber\\
\frac{d\sigma_{\text{diff ex}}}{d^2 b}= 
V_T &\equiv& \langle T^2 \rangle -  \langle T \rangle^2
\approx  \langle T^2 \rangle =\\
&=& A\int (1- e^{-F})^2 F^{-p} dF = (1-\frac{1}{2^{(2-p)}})\times 2\langle T \rangle.\nonumber
\label{eq:vtdis}
\end{eqnarray}
From these results we note that the ratio $d\sigma_{\text{diff ex}}/d\sigma_{\text{tot}} 
= V_T/(2\langle T \rangle)$ depends only on the value of the parameter $p$. As we have found
that $p$ is independent of the impact parameter for fixed $W$ and $Q^2$, we can
integrate over $b$, and find
\begin{equation}
\frac{\sigma_{\text{diff ex}}}{\sigma_{\text{tot}}} =  \frac{V_T}{2\langle T \rangle} \approx 1-\frac{1}{2^{2-p}}
\end{equation}
Thus the parametrisation in eq. (\ref{eq:powerdistr}) gives $ \sigma_{\text{diff ex}}/\sigma_{\text{tot}}\sim 0.18$ for $Q^2 = 14\,\mathrm{GeV}^2$ falling 
to $\sim 0.13$ at $Q^2 = 50\,\mathrm{GeV}^2$. Although the simple parametrisation overestimates the result of the MC, it gives a qualitatively correct result.

For a virtual photon in DIS the fluctuations will be further enhanced
by adding the fluctuations in the photon wave function, but this will not 
alter the conclusions presented above.

\subsection{$pp$ scattering}

The corresponding Born amplitude distributions in $pp$ collisions
are shown in fig. \ref{fig:f-distpp} for $W=100$ and $W=2000$ GeV and different $b$-values.
We note that here the interaction probability is large, which implies large
saturation effects.
The distributions can be well approximated by Gamma functions of the form
\begin{equation}
P(F) = A\, F^p\, e^{-aF}.
\label{eq:gammadistr}
\end{equation}
The distributions have two parameters, $p$ and $a$, which are tuned to the
MC results for different values of energy and impact parameter. The parameter 
$A$ is then fixed by the normalisation condition.
The result of the fit is shown in table \ref{tab:PFpar}, and we note here
that $a$ is essentially independent of the impact 
parameter, but falling with energy.
For a fixed energy, the decrease in $\langle F \rangle$ for more
peripheral collisions is related to a decrease in $p$ for 
larger $b$-values, pushing the distribution to smaller values of $F$. For fix $b$ the
parameter $p$ also grows with increasing energy, reflecting the larger interaction
probability.

\FIGURE[t]{\includegraphics[angle=270, scale=0.59]{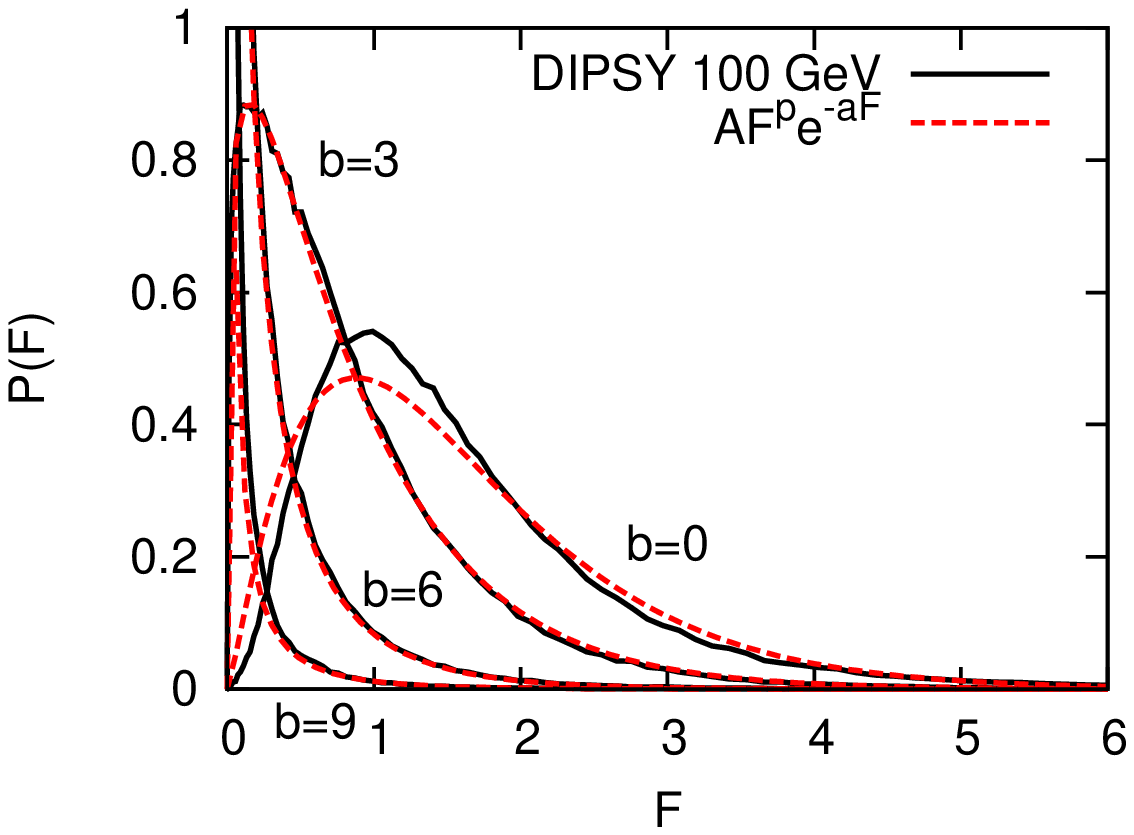}
\includegraphics[angle=270, scale=0.59]{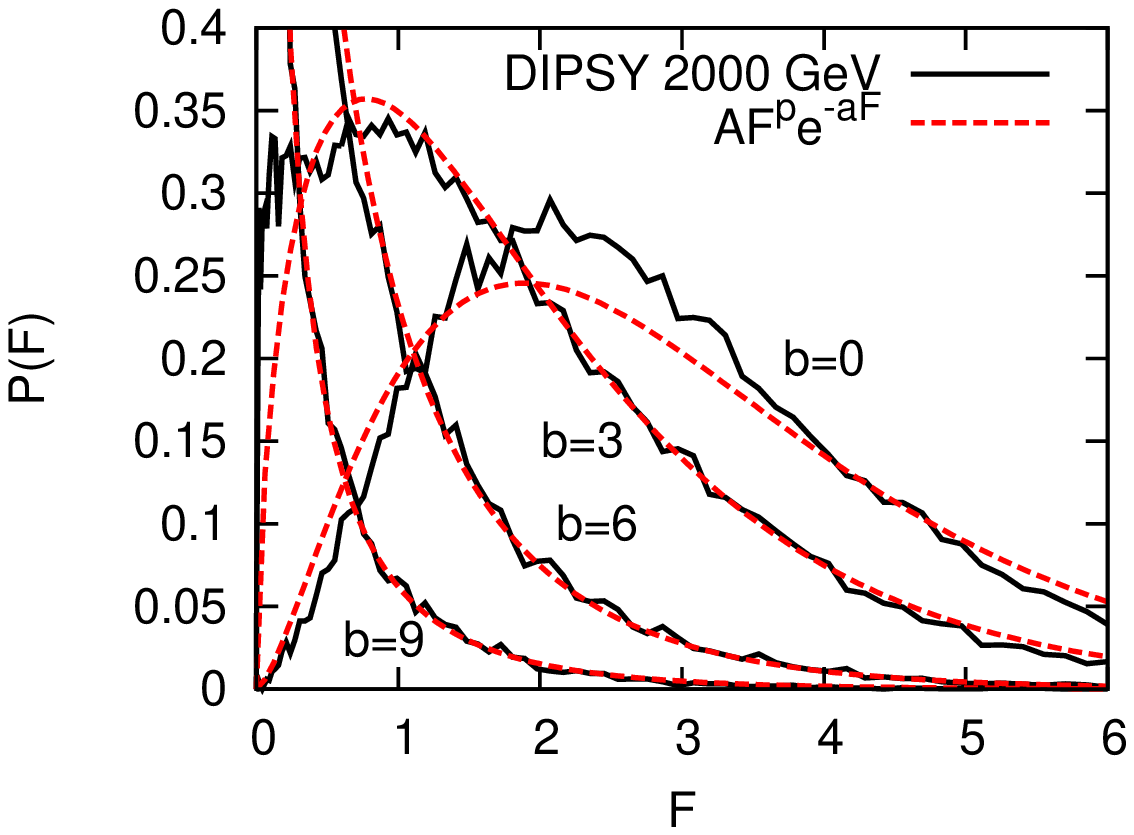}
   \caption{\label{fig:f-distpp}Probability distribution, $P(F)$, for the one-pomeron
amplitude $F$ in $pp$ collisions for $W=100$ $\mathrm{GeV}$
(left) and $W=2000$ $\mathrm{GeV}$ (right). $b$ is in units of GeV$^{-1}$. The dotted lines are fits of the form in eq. (\ref{eq:gammadistr}). } }

The parametrisation in eq. (\ref{eq:gammadistr}) gives
\begin{eqnarray}
\langle F \rangle& =&\frac{p+1}{a} \nonumber \\
\frac{V_F}{2\langle F \rangle}& =& \frac{1}{2a} \sim 0.35\,\,\, \mathrm{for} 
\,\,\,W=100\,\,\,\mathrm{GeV},
\end{eqnarray}
where $V_F \equiv \langle F² \rangle - \langle F \rangle²$ is the variance of $F$. Thus we find also here that the ratio 
between the variance and the average of the Born amplitude is independent of 
$b$. We note that this ratio is large, and similar to the result for 
$\gamma^* p$ collisions at lower $Q^2$-values. Thus, without saturation
we would have a correspondingly large value for 
$d\sigma_{\text{diff ex}}/d\sigma_{\text{tot}}$ also in $pp$ collisions.

\FIGURE[t]{\includegraphics[angle=270, scale=0.59]{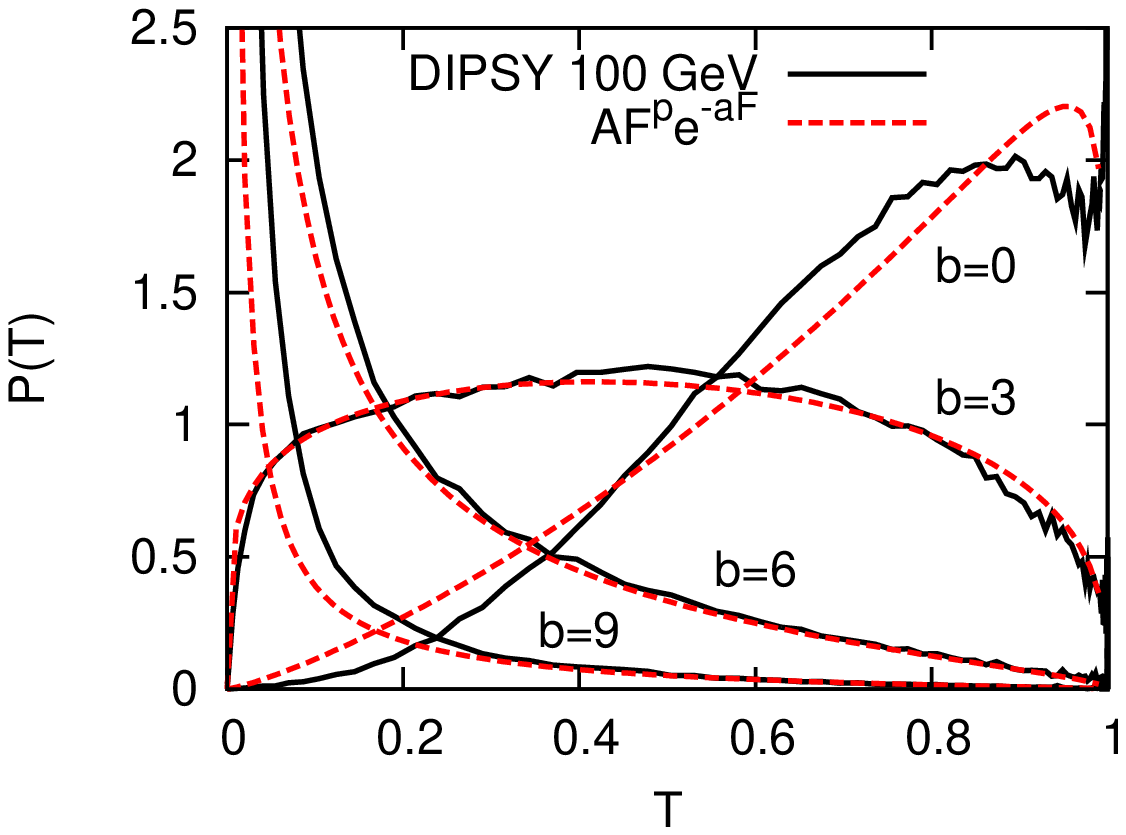}
\includegraphics[angle=270, scale=0.59]{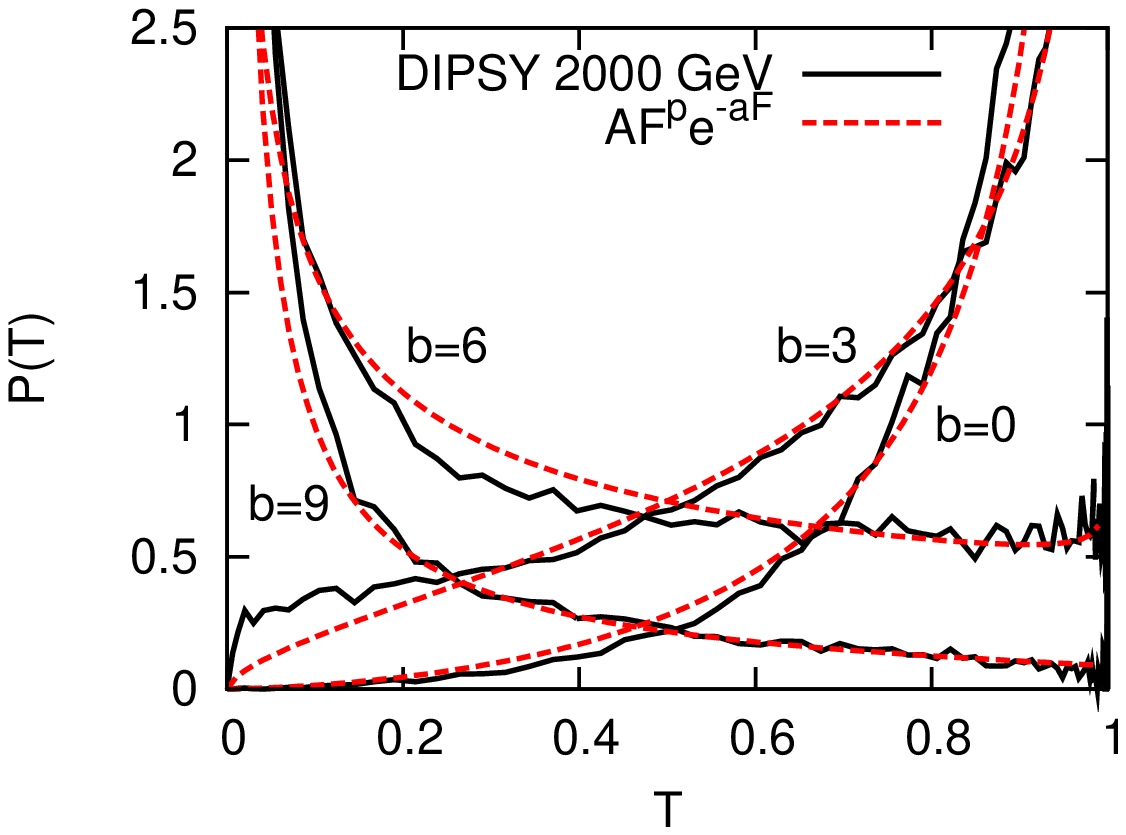}
   \caption{\label{fig:t-distpp}Probability distribution, $P(T)$, for the full amplitude $T$
for $pp$ collisions at $W=100$ GeV (left) and $W=2000$ GeV (right). $b$ is in units of GeV$^{-1}$. The dotted lines correspond to fits of the form in eq. (\ref{eq:gammadistr}).  } }

However, as the one pomeron amplitude $\langle F \rangle$ is large in $pp$ 
scattering, unitarity corrections are very
important. The probability distribution for the unitarised amplitude, $P(T;b)$, with $T=1-e^{-F}$, is shown in 
fig. \ref{fig:t-distpp} for $W=$ 100 and 2000 GeV and different $b$-values. We see 
that for the central collisions the distributions are very peaked close
to the unitary limit $T=1$. This reduces the fluctuations very strongly. 
For the parametrisation in 
eq. (\ref{eq:gammadistr}) the average and the variance
for the distribution in $T$ are also easily calculated, and given by
\begin{eqnarray}
\langle T \rangle& =& 1-(\frac{a}{a+1})^{p+1}\nonumber \\
V_T& =&(\frac{a}{a+2})^{p+1} - (\frac{a}{a+1})^{2p+2}.
\end{eqnarray}

We see that at high energies and central collisions, where the Born amplitude 
$\langle F \rangle$, and thus also the parameter $p$, become large, 
$\langle T \rangle$ will approach 1 and $V_T$ will go towards 0. For central collision at $W = 100$~GeV the ratio of diffractive events $V_T/(2\langle T \rangle)$ is about 0.035, a factor 10 lower than without unitarisation.
Therefore in central collisions diffractive excitation is suppressed, and
diffractive scattering is dominantly elastic.

The large effect of saturation in $pp$ collisions has also the effect that factorization 
is broken when comparing diffractive excitation in DIS and $pp$ collisions
\cite{Schilling:2002tz}.

\section{Impact parameter profile and $t$-dependence in $pp$-collisions}

In the previous section we showed the amplitude fluctuations for different impact parameter values.
We will here study the $b$-dependence, and the corresponding $t$-dependence, in more detail.
As mentioned above, diffractive excitation is small in central $pp$ collisions as $\langle T \rangle$ is approaching 1. In highly peripheral collisions 
both $\langle T \rangle$ and $V_T$ are small, again giving little diffractive excitation.
Therefore diffractive excitation is dominated by moderately peripheral
collisions, where $\langle T \rangle \sim 0.5$ and $\langle F \rangle \sim 1$.
The $b$-dependence of $\sigma_{\text{tot}}/2$, $\sigma_{\text{el}}$, and $\sigma_{\text{diff ex}}$ in the MC
is shown in fig. \ref{fig:b-distpp}.

\FIGURE[t]{\includegraphics[angle=270, scale=0.55]
{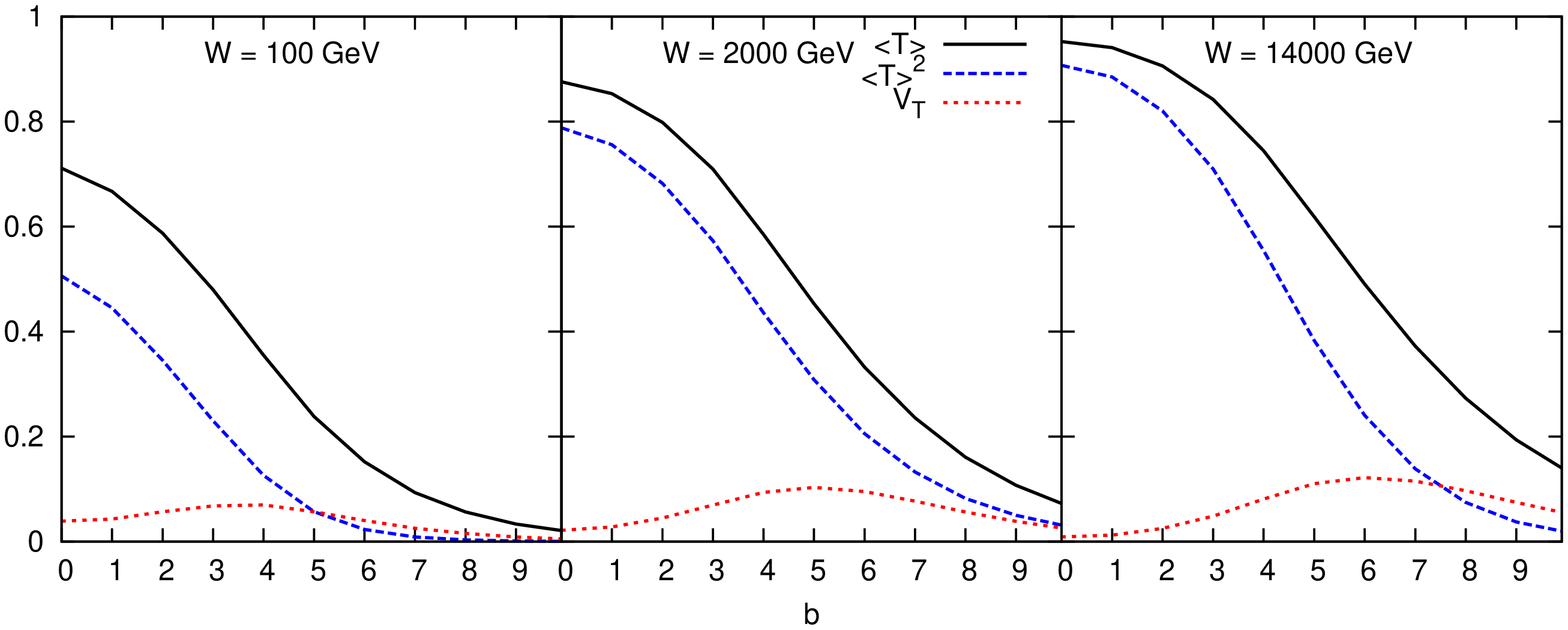}
   \caption{\label{fig:b-distpp}Impact parameter distributions from the MC for 
    $\langle T\rangle=(d\sigma_{\text{tot}}/d^2b)/2$, $\langle T\rangle^2=d\sigma_{\text{el}}/d^2b$, and 
    $V_T=d\sigma_{\text{diff ex}}/d^2b$ in
    $pp$ collisions at $W=100$, 2000, and 14000 GeV. $b$ is in units of GeV$^{-1}$. } }

As pointed out also in earlier analyses (\textit{e.g.} in refs.~
\cite{Miettinen:1978jb, Sapeta:2005ba}), this implies that diffractive 
excitation in $pp$ collisions appears in a ring with a radius which grows 
slowly with energy. In a purely perturbative calculation with massless
gluons, the total cross section will grow very fast due to the formation of
very large dipoles, and eventually violate Froisart's bound.  
However, as demonstrated by Avsar \cite{Avsar:2008dn}, the inclusion of 
confinement effects via a massive gluon (as in the simulations described above)
implies that very large dipoles are suppressed, and the black disk radius 
grows proportional to $\ln s$. 
This means that the total and elastic cross sections grow like $\ln^2 s$
for very large energies.
In addition the results in \cite{Avsar:2008dn} show that
the slope, when the interaction drops from black in central to white for more
peripheral collisions, is approximately constant with energy. 
Thus the width of the ring with large diffractive excitation is approximately constant at high energies.
Consequently  
the cross section for diffractive excitation will for very large energies
grow proportional to the radius of the ring, i.e. proportional to $\ln s$.

In our model the interaction is driven by absorption into inelastic channels, and with our definition,
where $S\equiv 1-T$, the imaginary part of the amplitude $T$ is neglected. 
It is therefore straight forward
to take the Fourier transform and calculate the $t$-dependence.
The result for the differential elastic $pp$ cross section was presented
in ref.~\cite{Flensburg:2008ag}, and the result for single diffractive excitation
is shown in fig.~\ref{fig:t-distsat}. Figure \ref{fig:t-distsat}$a$ shows the result at 
546 GeV, together with an extrapolation of a fit to UA8 data \cite{Bruni:1993ym}, normalised to
the model result. The UA8 data cover only the range $0.8<|t|<2.5\,\text{GeV}^2$,
but we note that the $t$-slope in the model agrees well 
with this fit. For comparison also the elastic cross section 
is included in the figure, and we see that the diffractive slope is 
significantly smaller
than the slope in elastic scattering. This is a consequence of the larger
$b$-values for diffractive excitation shown in fig.~\ref{fig:b-distpp}.

Figure ~\ref{fig:t-distsat}$b$ shows how the differential 
cross section for diffractive excitation is varying with energy.
We see that the increase is quite slow, as a result of saturation and
unitarity constraints. Note that for low $t$, the energy dependence is
stronger, due to the growth of the radius of the ring, while the energy dependence for high $t$ is much slower, showing that the width of the ring is almost energy independent, in agreement with 
ref.~\cite{Avsar:2008dn}.

\FIGURE[t]{\includegraphics[angle=270, scale=0.59]{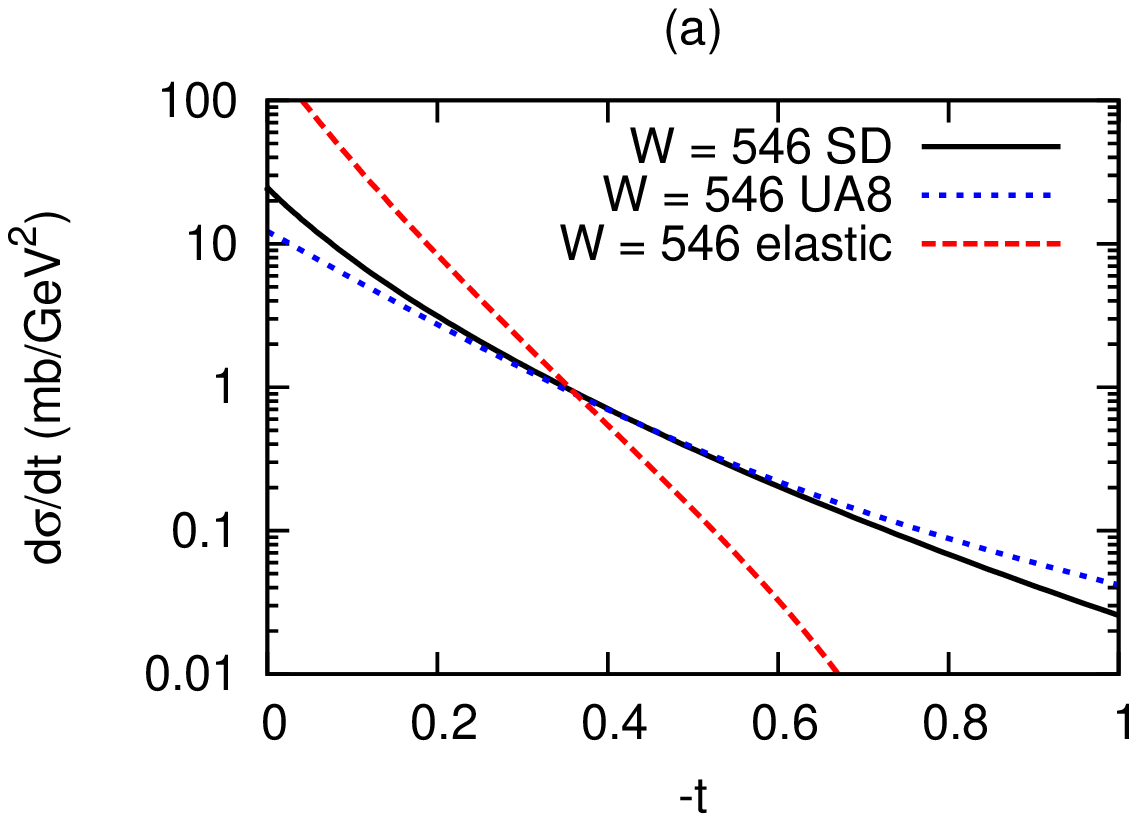}
\includegraphics[angle=270, scale=0.59]{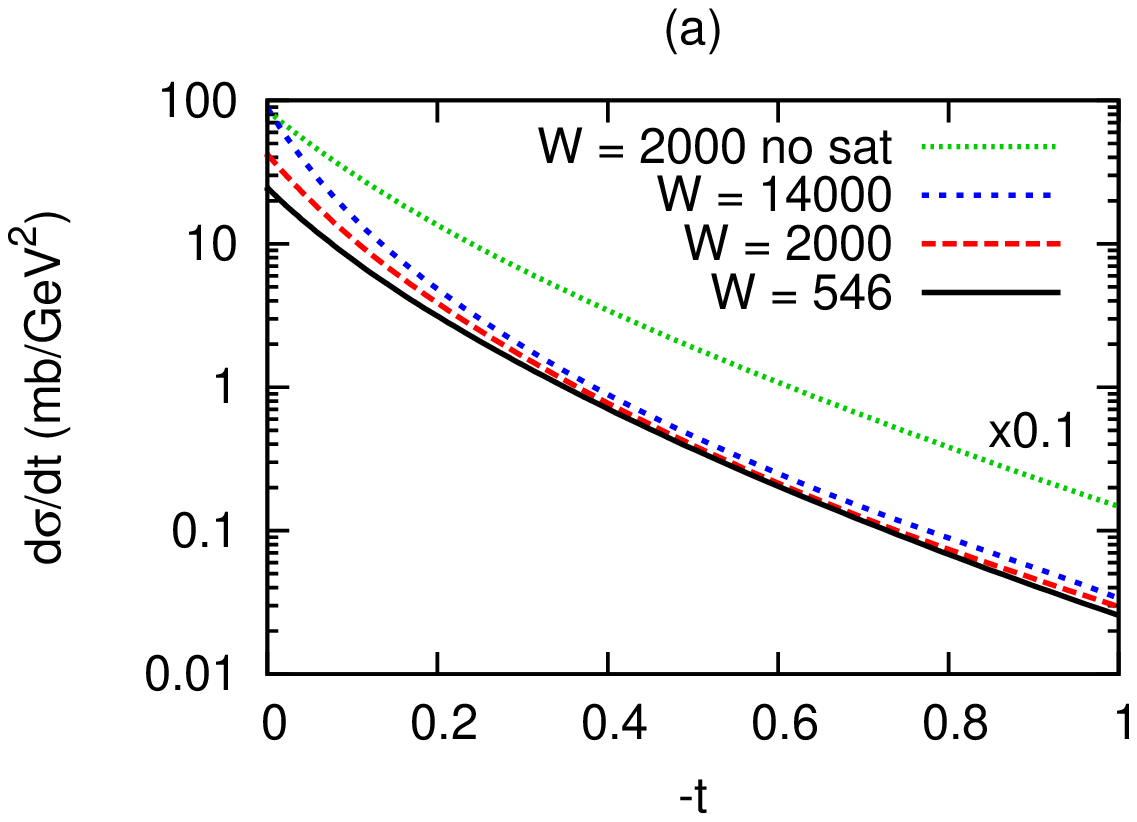}
  \caption{\label{fig:t-distsat}(a) $t$-dependence for the single diffractive cross section
    at 546 GeV from the MC, together with a fit to UA8 data. The elastic cross section
    is included for comparison. (b) MC results for the
    energy dependence of the $t$-distribution for single diffractive excitation. The effect
    of saturation is demonstrated by the dotted
    line, which shows the result at 2000 GeV without saturation, scaled by 
    a factor 0.1. } }

To further illustrate the effect of saturation we show in 
fig.~\ref{fig:t-distsat}$b$ also the $t$-dependence of the single diffractive cross section
obtained from the unsaturated Born amplitude. We see that saturation
reduces the cross section by roughly a factor 25 at 2000 GeV.  The slope is 
less affected, but the suppression for small $b$-values implies that the $t$-dependence 
deviates more from a pure exponential, when saturation is included. 

\section{Relation Good--Walker -- Triple-Regge}

In this section we will discuss the relation between the results using the
Good--Walker formalism described above, and the triple-Regge formalism.
In this comparison we want to study the contribution from the bare pomeron,
meaning the one-pomeron amplitude without contributions from saturation,
enhanced diagrams or gap survival form factors. We want to see if the 
fluctuations in the dipole cascades reproduce the powerlike energy
dependence expected in the Regge formalism.

When $s$, $M_{\mathrm{X}}^2$, and $s/M_{\mathrm{X}}^2$ are not small, pomeron exchange should 
dominate. If the pomeron is a simple pole we expect the following 
expressions for the $pp$ total and diffractive cross sections:
\begin{eqnarray}
\si_{\text{tot}}&=&\be^2(0)s^{\al(0)-1} \equiv \si_0^{p\bar{p}}s^\ep,\nonumber\\
\frac{d\si_{\text{el}}}{dt} &=& \frac{1}{16\pi}\be^4(t)s^{2(\al(t)-1)}, \nonumber\\
M_{\mathrm{X}}^2 \frac{d\si_{\text{SD}}}{dtd(M_{\mathrm{X}}^2)} &=& \frac{1}{16\pi}\be^2(t)\be(0)g_{3\text{P}}(t)
\left( \frac{s}{M_{\mathrm{X}}^2} \right)^{2(\al(t)-1)}\left( M_{\mathrm{X}}^2 \right)^{\epsilon}.
\label{eq:barepomeron}
\end{eqnarray}
Here $\al(t)=1+\ep+\al ' t$ is the 
pomeron trajectory, and $\beta(t)$ and $g_{3P}(t)$ are the proton-pomeron and
triple-pomeron couplings respectively. (We have here omitted the scale 
$s_0$ in the powers $(s/s_0)^\alpha$ or $(M_{\mathrm{X}}^2/s_0)^\alpha$. This scale is in 
the following is assumed to be 1 $\mathrm{GeV}^2$.)

\FIGURE[t]{\includegraphics[width=0.5\linewidth,angle=270]{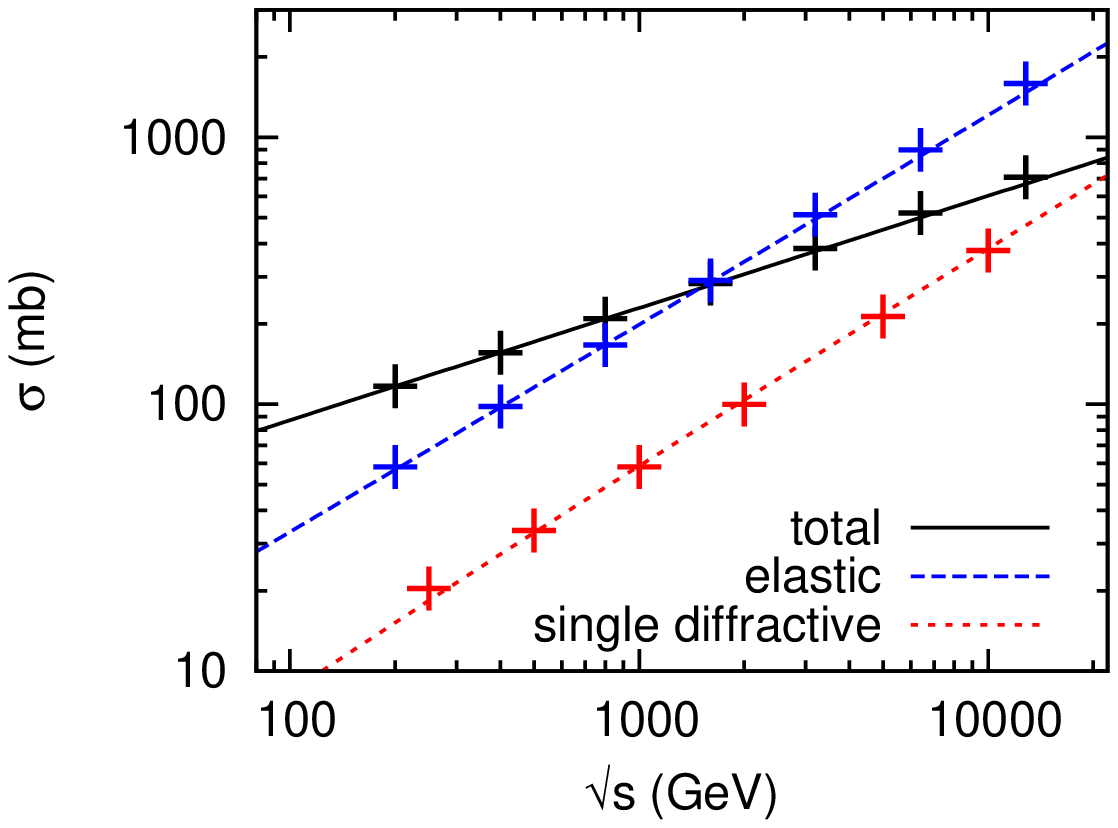}
  \caption{\label{fig:ppSD}The total, elastic and single diffractive cross
    sections in the \emph{one-pomeron} approximation. The crosses are from the 
    dipole cascade model without saturation,
    and the lines are from a tuned triple Regge parametrisation.}}

The results of the MC for the total, elastic, and single diffractive 
cross sections are shown by the crosses in fig.~\ref{fig:ppSD}. 
The elastic and diffractive cross 
sections are integrated over $t$ and $M_{\mathrm{X}}^2$. The single diffractive cross
section is calculated in the
total cms, which corresponds to an integration over masses
in the range $M_{\mathrm{X}}^2<\sqrt{s}\cdot 1$ GeV, and it corresponds to
excitation of one side only. We see that the result indeed has the
powerlike increase with energy, which is characteristic for a Regge pole. We
also note that in the one-pomeron approximation the elastic cross section is
larger than the total for $\sqrt{s} > 15\,$GeV.

If we assume a simple exponential form for the proton-pomeron coupling, 
$\beta(t)= \beta(0)\exp(b_{0,\text{el}}\, t/4)$, we can integrate the
elastic cross section in eq.~(\ref{eq:barepomeron}) over $t$, and obtain
\begin{equation}
\sigma_{\text{el}}=\frac{\si_{\text{tot}}^2}{16\pi B(s)}, \,\,\,\,\,
\mathrm{with}\,\,\, B(s)=b_{0,\text{el}}+2\al' \ln s.
\label{eq:integratedelastic}
\end{equation}
Besides the shrinking of the elastic peak, the pomeron slope $\alpha'$ 
also gives a logarithmic correction to the powerlike 
increase of the elastic cross section. A consistent fit to both quantities is
obtained for $\alpha'=0.2\,\,\text{GeV}^{-2}$. In fig. \ref{fig:ppSD} the lines are obtained from 
the expressions in eqs.~(\ref{eq:barepomeron}, \ref{eq:integratedelastic}) with 
the parameter values
\begin{eqnarray}
\al(0)&=&1+\epsilon =1.21, \,\,\,\,\al' = 0.2\,\text{GeV}^{-2},\nonumber\\
\si_0^{p\bar{p}}&=&\beta^2(0)=12.6\,\text{mb}, \,\,\,\, 
b_{0,\text{el}} = 8\,\text{GeV}^{-2}, \,\,\,\, g_{3\text{P}}(t) =\text{const.} = 0.3\,\text{GeV}^{-1}.
\label{eq:parameters}
\end{eqnarray}
We have here assumed a constant triple-pomeron coupling, and we see that 
the MC results in fig. \ref{fig:ppSD} are very well reproduced by this fit.

\FIGURE[t]{\includegraphics[angle=270, scale=0.8]{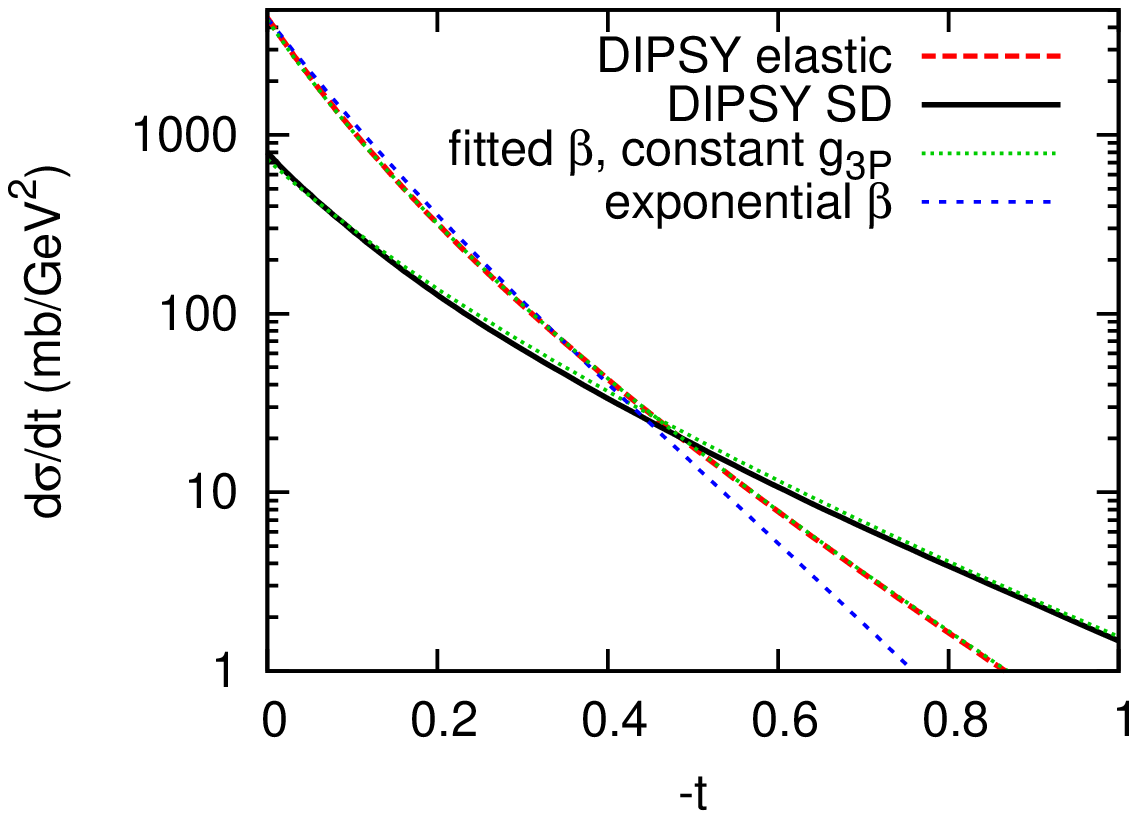}
  \caption{\label{fig:t-distnosat}$t$-dependence for elastic and single 
    diffractive
    excitation without saturation effects, at 1800 GeV. A very close fit to 
    $d\si_{\text{el}}/dt$ is obtained from the proton-pomeron coupling in
    eq.~(\ref{eq:betat}). Including a constant triple-pomeron coupling 
    in the expression for diffractive excitation gives the thin dotted line,
    which gives a good description of the model result.
    For comparison also a pure exponential fit to the elastic cross section is included.} }

The $t$-dependence of elastic scattering and diffractive excitation, shown
in fig.~\ref{fig:t-distnosat}, are however not pure exponentials, as assumed
in the fit above. (For diffractive excitation a minor deviation from a pure 
exponent originates from the integration over $M_X^2$.) We want to study this 
dependence in some more detail, and are
here in particular interested in the $t$-dependence of the triple-pomeron
coupling $g_{3P}(t)$. A very close fit to $d\si_{\text{el}}/dt$
is obtained for (with $t$ measured in $\text{GeV}^2$)
\begin{equation}
\beta^4(t)=\beta^4(0)\exp\left(\frac{10\,t}{1-1.8\,t}\right),
\label{eq:betat}
\end{equation}
which cannot be distinguished from the MC result in fig.~\ref{fig:t-distnosat}.
Inserting this fit into the expression for the diffractive cross section
integrated over $M_{\mathrm{X}}^2$, and assuming a constant triple-pomeron coupling
equal to $0.27 \,\text{GeV}^{-2}$,
gives the thin dotted line in fig.~\ref{fig:t-distnosat}. We see that this is
quite a good fit. It is slightly less steep for $|t|$-values below  $0.1
\,\text{GeV}^2$, but deviating less than 10\% from the result of the model. We
also note that these modifications of the $t$-dependence for the elastic and
diffractive cross sections do not modify the good fits to the integrated cross
sections in fig. \ref{fig:ppSD}.
\vspace{1mm}

We here want to make the following comments:

\begin{itemize}
\item
 \emph{Comparison with perturbative QCD}

The expressions in eq.~(\ref{eq:barepomeron}) correspond to a pomeron which is
a simple pole. This is not the case in perturbative QCD. In the LL approximation 
the pomeron is a cut in the angular momentum plane, which gives 
logarithmic corrections
to the proton-pomeron coupling: $\beta(0)\sim 1/(\ln s)^{1/4}$ while 
$\beta(t)\sim 1/(\ln s)^{3/4}$ for $t\neq 0$ \cite{Mueller:1994jq, Bartels:2002au}.
The different $s$-dependence when $t$ is equal to, or different from, zero is
associated with a cusp in the $t$-dependence at $t=0$. However, at these small
$t$-values perturbative QCD is not applicable, and non-perturbative effects are
important. It was also pointed out by Lipatov \cite{Lipatov:1985uk}, that a 
running coupling can modify the cut to a series of poles. (Note that a running 
coupling is included in our model simulations.)
A strong increase in the elastic cross section at very small $t$-values 
is not seen in the experimental data, and also not present in our result
in fig.~\ref{fig:t-distnosat}. An extra factor of $1/(\ln s)^{1/4}$ in
$\beta(0)$ and $\beta(t)$ would give an equally good fit to the results in
fig.~\ref{fig:ppSD}, provided the pomeron intercept is
increased to $\alpha(0)=1.25$. It is, however, not possible to find a good
fit if $\beta(t)$ is proportional to $1/(\ln s)^{3/4}$, inside a $t$-region
essential for the integrated elastic cross section.

In LL perturbative QCD also the triple-pomeron coupling has a singular 
behaviour at $t=0$ \cite{Mueller:1994jq, Bartels:2002au}:
\begin{equation}
g_{3\text{P}}(t)\sim \frac{1}{(\ln M_X^2)^{1/4} (\ln s/M_X^2)^{3/4}} \frac{1}{\sqrt{-t}}
\end{equation}
We saw above that our results were
well reproduced by a constant triple-pomeron coupling, only slightly
underestimating the slope for small $t$-values below $|t|=0.1$, and our 
analysis does not support
a triple-pomeron coupling with a strong $t$-dependence. A slowly varying 
triple-pomeron coupling is also in agreement 
with early analyses \cite{Kaidalov:1973tc, Field:1974fg}.
We note, however, that the magnitude of the triple-pomeron coupling agrees 
(within the large uncertainties) with the perturbative estimate in 
ref.~\cite{Bartels:2002au}, which in our notation corresponds to
$\pi g_{3P} \sim 0.2 - 1.7 \text{GeV}^{-1}$ for $0.25<|t|<4 \,\text{GeV}^2$.

\item
 \emph{Comparison with other analyses}

We should also note that the bare pomeron is not an observable. Here we have
included NLL effects and confinement in the evolution, but not nonlinear 
effects from saturation or multiple collisions. The effect of nonlinearity
will appear differently if one first adds NLL effects and then compares
the results with and without saturation, as compared with an approach where
the LL result is compared with and without saturation, before NLL effects
are included. Therefore the bare pomeron may look different, depending
upon the scheme used to remove the nonlinear effects.

Keeping this in mind, we want to compare our result in 
eq.~(\ref{eq:parameters}) with some recent more
traditional Regge analyses. As examples the bare pomeron in the analysis by
Ryskin \emph{et al.} \cite{Ryskin:2009tj} has three components with different dependence on
the impact parameter, or $t$, in order to mimic the branch cut structure
of the pomeron singularity. The three poles have the same intercept equal to
$\alpha(0) =1.3$, and quite small slopes. The dominant component has
$\alpha'=0.05\,\,\mathrm{GeV}^{-2}$, and the slope of the other components
are even smaller. 
Ostapchenko \cite{Ostapchenko:2010gt} finds in an approach with two
pomerons $\alpha(0)\approx 1.35, \alpha'\approx 0.08 \mathrm{GeV}^{-2}$
and $\alpha(0)\approx 1.15, \alpha'\approx 0.14 \mathrm{GeV}^{-2}$
for the hard and soft pomerons respectively.
Gotsman \emph{et al.} \cite{Gotsman:2008tr} find in an analysis with a single pomeron $\alpha(0)=1.335$ and $\alpha'=0.01 \mathrm{GeV}^{-2}$.
In another fit with a single pomeron pole Kaidalov \emph{et al.} \cite{Kaidalov:2009hn} 
find $\alpha(0) =1.12$ and $\alpha'=0.22\,\,\mathrm{GeV}^{-2}$. We can also
compare with the results by Goulianos \cite{Goulianos:1995vn,
  Goulianos:2004as}, who in a formalism with a renormalised pomeron flux finds
the values $\alpha(0) 
\approx 1.11$ and $\alpha'=0.26\,\,\mathrm{GeV}^{-1}$. We see here that the
way saturation is taken into account can have a large effect on the result
for the bare pomeron. We also note that our result lies somewhere in between
these examples.
\end{itemize} 
\vspace{1mm}

In conclusion we see that the Born amplitude in our dipole cascade model 
indeed reproduces the triple-Regge formula for a bare pomeron pole with 
$\alpha(0)=1.21$, $\alpha'=0.2$, and an almost constant triple-pomeron
coupling. The $t$-dependence of both the elastic and the diffractive cross
sections is close to an exponential. There is no indication for more dramatic 
variations, like those obtained in LL perturbative QCD, where the pomeron is a
cut singularity in the angular momentum plane. Here one also expects 
$\beta(0)\sim1/(\ln s)^{1/4}$ while $\beta(t)\sim1/(\ln s)^{3/4}$ for $t\neq0$.
Our result would be consistent with a proton-pomeron coupling proportional to
$1/(\ln s)^{1/4}$, if the intercept is increased to 1.25, but not with
$\beta(t)\sim1/(\ln s)^{3/4}$.

\section{Conclusions}

Diffractive excitation represents a large fraction of the cross section in
$pp$ collisions or DIS. In the Good--Walker formalism diffractive excitation 
is determined by the fluctuations in the scattering amplitude. In traditional 
applications this formalism has been limited to small mass excitation, while
excitation to high masses has been described in the triple-Regge formalism,
introducing a set of parameters for the pomeron couplings. It was 
demonstrated by Mueller and Salam \cite{Mueller:1996te} that BFKL evolution
contains large fluctuations, and in this paper we demonstrate that, by
including these fluctuations in the analysis, it is possible to describe
diffractive excitation to both low and high masses. 

The Lund Dipole Cascade model, implemented in the DIPSY MC, describes
successfully the total, elastic, and diffractive cross sections in DIS and $pp$
collisions \cite{Avsar:2005iz, Avsar:2006jy, Flensburg:2008ag}. 
In this paper we show how the fluctuations in the BFKL evolution
can reproduce diffractive excitation to low and high masses within 
the Good--Walker formalism, with parameters determined only from the total
and elastic cross sections. In DIS at HERA the fluctuations give a diffractive 
cross section of the order of 10\%, and saturation has a relatively small
effect on the result. However, in $pp$ collisions unitarity constraints and 
saturation reduce the fluctuations, when the scattering approaches the black
disc limit. Therefore diffractive excitation in high energy $pp$ collisions is 
dominated by peripheral collisions, and pomeron exchange in DIS and $pp$ 
collisions does not factorise. 

In the triple-Regge formalism, saturation effects are included in terms of
``enhanced diagrams'', gap-survival form factors or saturation effects in the
pomeron flux, which reduce the effect of the ``bare pomeron''. In this paper
we have also studied the effective bare pomeron, corresponding to the result 
obtained when non-linear effects are not included. We see that the result 
indeed is well described by a bare pomeron pole, with $\alpha(0)=1.21$ and
$\alpha' = 0.2\, \mathrm{GeV}^{-2}$. The triple-pomeron coupling is fixed by the couplings in the BFKL ladder, and the results are well reproduced by a constant coupling $g_{3\text{P}} \approx 0.3\, \mathrm{GeV}^{-1}$. Although our model
is based on perturbative QCD, with non-perturbative effects introduced
only in the incoming proton wavefunction and confinement effects via an 
effective gluon mass, this result contrasts to LL BFKL, where the pomeron is a
cut singularity and the pomeron couplings have strong $t$-dependencies.

In this paper we have not discussed the properties of exclusive final states
in diffractive excitation (apart from the mass distribution). We hope to 
return to this problem and to hard diffraction in future work.

The main results can be summarised as follows:

- The \emph{fluctuations} in a BFKL ladder are large. Taking these 
fluctuations into account as in the DIPSY MC, it is possible to describe 
diffractive excitations to both low and
high masses within the Good--Walker formalism.

- \emph{Saturation} effects are small in DIS, but large in $pp$ collisions. 
Therefore diffractive excitation is a peripheral process in $pp$ scattering, 
and factorisation of pomeron exchange is broken.

- The result of the Good--Walker formalism \emph{reproduces the triple-Regge}
result for diffractive excitation, with a bare pomeron pole with 
$\alpha(0)=1.21$, $\alpha' = 0.2\, \mathrm{GeV}^{-2}$, and an almost 
constant triple-pomeron coupling $g_{3P} \approx 0.3\, \mathrm{GeV}^{-1}$.

\section{Acknowledgements}

We want to thank Leif L\"onnblad for valuable discussions. Work supported in part by the Marie
Curie RTN ``MCnet'' (contract number MRTN-CT-2006-035606).

\bibliographystyle{utcaps}
\bibliography{/home/william/people/leif/personal/lib/tex/bib/references,refs}

\end{document}